\documentclass[prd,aps,notitlepage,superscriptaddress,nofootinbib]{revtex4-1}
\usepackage{amsmath}
\usepackage{epsfig}
\def\bea#1\eea{\begin{align}#1\end{align}} 
\newcommand{\nnu}{\nonumber\\}
\newcommand{\bef}{\begin{figure}[htb]\centering}
\newcommand{\eef}{\end{figure}}
\allowdisplaybreaks

\begin{document}
\title{Next-to-leading order transverse momentum broadening \\ for Drell-Yan production in p+A collisions}

\date{\today}

\author{Zhong-Bo Kang}
\email{zkang@lanl.gov}
\affiliation{Theoretical Division, 
                   Los Alamos National Laboratory, 
                   Los Alamos, NM 87545, USA}

\author{Jian-Wei Qiu}
\email{jqiu@bnl.gov}
\affiliation{Physics Department, Brookhaven National Laboratory, Upton, NY 11973, USA}
\affiliation{C.N. Yang Institute for Theoretical Physics and Department of Physics and Astronomy, 
Stony Brook University, Stony Brook, NY 11794, USA}

\author{Xin-Nian Wang}
\email{xnwang@lbl.gov}
\affiliation{Institute of Particle Physics and Key Laboratory of Lepton and Quark Physics (MOE),
                  Central China Normal University, 
                  Wuhan 430079, China}
\affiliation{Nuclear Science Division, 
                   Lawrence Berkeley National Laboratory, 
                   Berkeley, CA 94720, USA}

\author{Hongxi Xing}
\email{hxing@lanl.gov}
\affiliation{Theoretical Division, 
                   Los Alamos National Laboratory, 
                   Los Alamos, NM 87545, USA}

\begin{abstract}
We present the nuclear transverse momentum broadening for Drell-Yan lepton pair production in p+A collisions at next-to-leading order (NLO) in powers of strong coupling constant $\alpha_s$.
We verify that the transverse momentum weighted differential cross section in NLO perturbative QCD (pQCD) at twist-4 can be factorized into the convolution of parton distribution function of an active parton in the projectile proton, a twist-4 multiparton correlation function of the target nucleus, and perturbatively calculable hard coefficient function. We identify a QCD evolution equation for such a twist-4 nuclear gluon-quark correlation function, and verify its universality -- its process independence.
This finding demonstrates the prediction power of pQCD factorization approach for studying parton multiple scattering in nuclear medium. 
\end{abstract}

\maketitle

\section{Introduction}
Transverse momentum broadening is commonly defined as the difference between the average transverse momentum square for the particles produced in nuclear collisions and that in hadronic collisions. Such a quantity plays an important role in understanding QCD dynamics of parton multiple scattering in the large nuclei, as well as in using hard processes to probe the properties of the nuclear medium~\cite{Qiu:2001hj,Abreu:2007kv,Boer:2011fh,Albacete:2013ei}. This observable has been studied extensively in theory for various processes in proton-nucleus (p+A) and lepton-nucleus ($e$+A) collisions~\cite{Guo:1998rd,Guo:2000eu,Luo:1993ui,Fries:2002mu,Qiu:2003pm,Kang:2008us,Kang:2011bp,Kang:2012am,Xing:2012ii}. It has also been measured in deeply inelastic scattering (DIS) at Jefferson Lab (JLab), HERA and p+A experiments at the Fermilab (FNAL), the Relativistic Heavy Ion Collider (RHIC) and the Large Hadron Collider (LHC). These measurements include hadron production in $e$+A collisions~\cite{Airapetian:2009jy,Brooks:2009xg}, dijet production in $\gamma/\pi$+A~\cite{Naples:1994uz} and p+A collisions~\cite{Adler:2005ad,Chatrchyan:2014hqa,Adam:2015xea}, as well as for vector boson (such as Drell-Yan (DY), $J/\psi$, and $\Upsilon$) production in p+A collisions~\cite{McGaughey:1999mq,Peng:1999gx,Alde:1991sw,Alde:1990im,Leitch:1995yc,Vasilev:1999fa,Johnson:2006wi,Adare:2012qf,Adam:2015jsa,Adamczyk:2016dhc}.

In this paper we study the transverse momentum broadening for Drell-Yan lepton pair production in p+A collisions. This is an ideal process to probe the QCD dynamics of initial-state parton multiple scattering since the final-state virtual photon $\gamma^*$ (or the decayed lepton pair $\ell^+\ell^-$) does not interact with the nuclear medium via strong interaction. This process is complementary to the single hadron production in semi-inclusive deep inelastic scattering (SIDIS), where the transverse momentum broadening of the observed hadron is sensitive to the final-state parton multiple scattering~\cite{Kang:2014ela}. Experimentally, the E772 and E866 collaborations at Fermilab have analyzed transverse momentum broadening of DY process in p+A collisions with various nuclear targets~\cite{McGaughey:1999mq,Peng:1999gx,Johnson:2006wi}, where 
consistent and relatively large transverse momentum broadening was observed. With the advent of new experimental opportunities from Fermilab-E906 \cite{E906}, RHIC and the LHC, transverse momentum broadening for DY production will be measured to wider kinematic regions at different center-of-mass energies. This will further illuminate the QCD dynamics for parton multiple scatterings, and provide new insights into the nuclear medium properties.

Previous calculations of DY transverse momentum broadening have been limited to the leading order (LO) in the expansion of the strong coupling constant~\cite{Guo:1998rd,Kang:2008us,Kang:2012am}. The initial-state parton double scattering in the large nucleus is the main source that generates the broadening and can be treated as power corrections within a generalized high-twist factorization formalism~\cite{Luo:1992fz,Luo:1994np,Qiu:1990xy}. However, a complete next-to-leading order (NLO) analysis is still lacking for high precision determination of the medium properties. Such a NLO analysis is the main focus of our current paper. We will follow the same high-twist factorization approach to perform the calculations, which has the advantage to readily quantify the radiative corrections in powers of strong coupling constant. For other different approaches in studying multiple scattering and the associated transverse momentum broadening, see, e.g., Refs.~\cite{Mueller:2012bn,Liou:2013qya,Blaizot:2014bha,Iancu:2014kga,Wu:2014nca}.

Within the high-twist formalism, we have previously computed the transverse momentum broadening for SIDIS process at the NLO, where we have verified the validity of the high-twist factorization for final-state double scattering inside a large nucleus at the one-loop order~\cite{Kang:2013raa,Xing:2014kpa}. In this paper, we extend our previous calculations for SIDIS to DY process. We will evaluate at the NLO  the initial-state double scattering contributions to the transverse momentum broadening for the lepton pair production in DY process in p+A collisions. We believe that this paper is a valuable addition to our previous work on SIDIS~\cite{Kang:2013raa,Kang:2014ela}, not only because of the great interest relevant to the experiments at FNAL, RHIC, and LHC facilities, but also the theoretical significance in further testing the QCD factorization at twist-4 level for initial-state double scattering, as well as demonstrating the universality of the associated high-twist multi-parton correlation functions~\cite{Kang:2013raa}. 

The rest of our paper is organized as follows. In Sec.~\ref{sec-LO}, we introduce our notations, and review the LO result for transverse momentum broadening. In Sec.~\ref{sec-NLO}, we present our NLO results for transverse momentum broadening by including both the gluon-quark and gluon-gluon double scattering, and will leave the calculation of the quark-quark double scattering to a future publication. We show explicitly how the soft divergences cancel out between real and virtual corrections, how the remaining collinear divergences can be absorbed into the corresponding parton distribution functions from the beam proton, and/or the twist-4 parton correlation functions of the colliding nucleus, as well as how the evolution equations of these correlation functions on the factorization scale can be derived. We conclude our paper in Sec.~\ref{sec-sum}.

\section{Transverse momentum broadening at leading order}
\label{sec-LO}

We consider Drell-Yan dilepton production in p+A collisions,
\bea
P(p')+A(p) \to [\gamma^*\to] \ell^+\ell^-(q)+X,
\eea
where $q$ is the four-momentum of the lepton pair (or virtual photon $\gamma^*$) with invariant mass $Q^2=q^2$, $p'$ is the momentum of the projectile proton, and $p$ is the momentum per nucleon of the target nucleus with atomic number $A$. The transverse momentum broadening of the lepton pair in DY process is defined as the difference between the averaged transverse momentum square of the observed dilepton produced in a p+A collision and that in a proton-proton (p+p) collision,
\bea
\Delta \langle q_{T}^2\rangle = \langle q_{T}^2\rangle_{pA} - \langle q_{T}^2\rangle_{pp},
\eea
where $q_T$ is the transverse momentum of the produced dilepton, with $\langle q_{T}^2\rangle$ defined as
\bea
\langle q_{T}^2\rangle = 
\int dq_{T}^2q_{T}^2 \left.\frac{d\sigma}{dQ^2dq_{T}^2}\right/\frac{d\sigma}{dQ^2}.
\label{eq-qt2}
\eea
\bef
\psfig{file=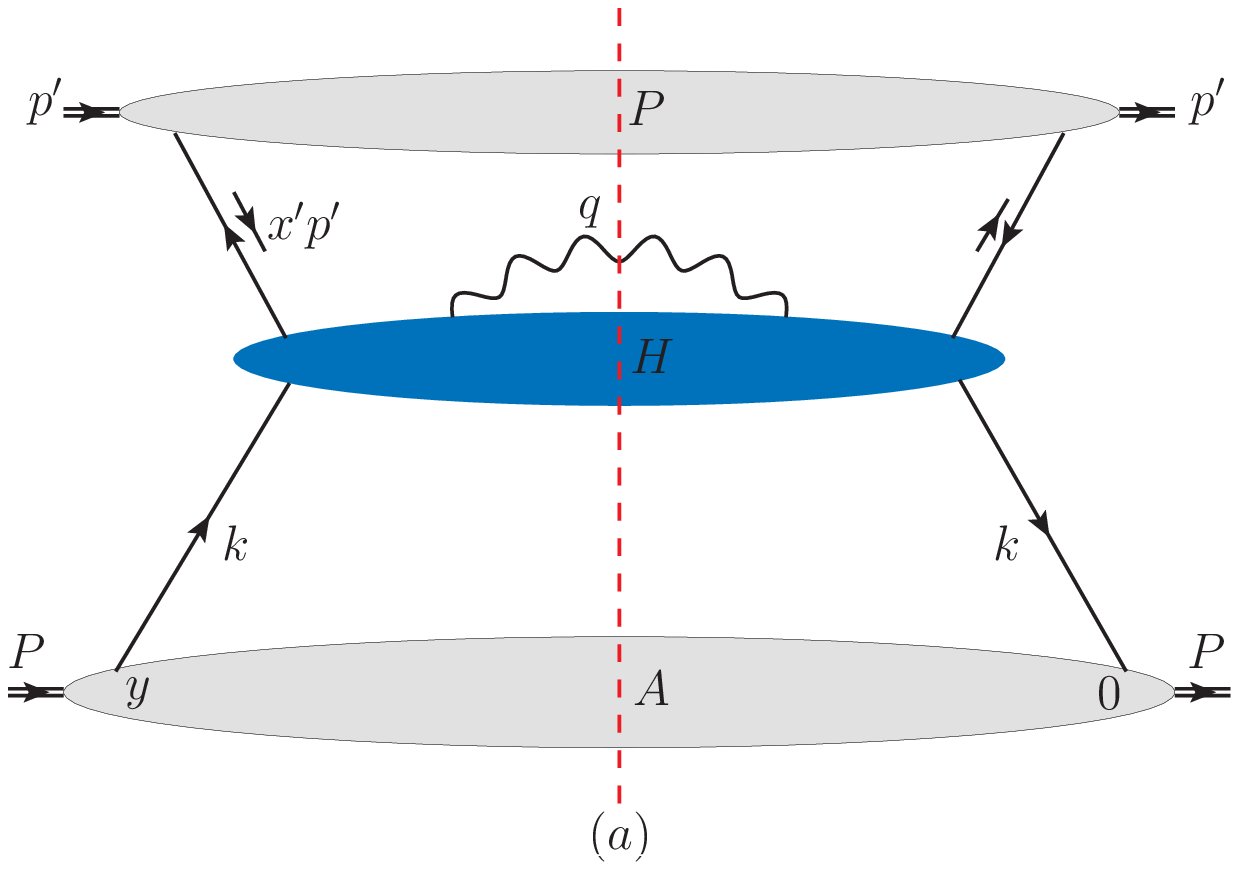, width=2.6in}
\hskip 0.4in
\psfig{file=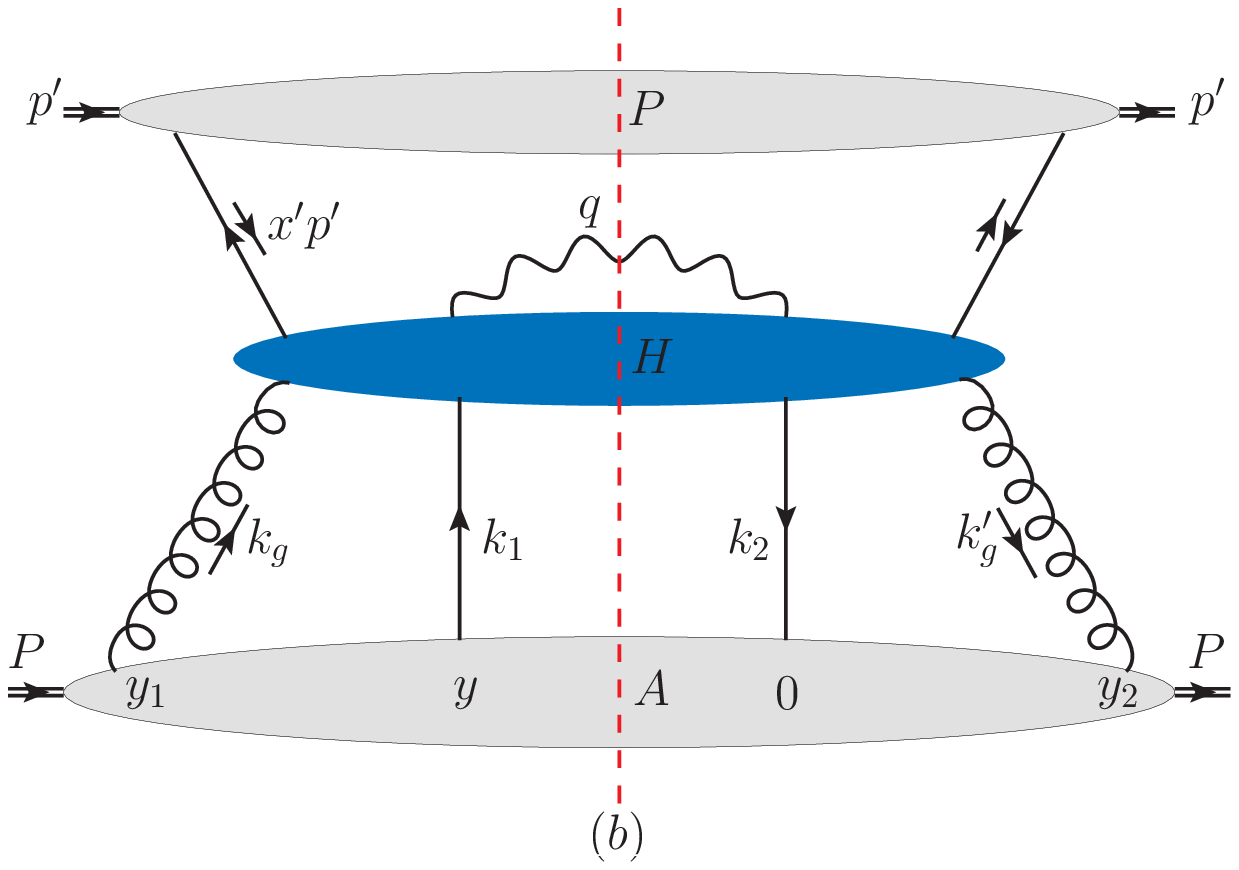, width=2.6in}
\caption{The general diagrams for Drell-Yan production in p+A collisions: (a) single scattering contribution with parton momentum $k=xp$ entering the partonic hard part $H$ (the blue blob on-line), (b) double scattering contribution with parton momenta $k_1=x_1p, ~k_2=(x_1+x_3)p+k_{2T} - k_{3T}, ~k_g = x_2p+k_{2T}, ~ k_g'=(x_2-x_3)p+k_{3T}$. Here $k_{2T}$ and $k_{3T}$ are the transverse momentum kicks from the nucleus. 
}
\label{fig:general}
\eef

Let us explore the physical picture behind the transverse momentum broadening for Drell-Yan production in p+A collisions. In the large target nucleus, the incoming parton (e.g., a quark) from the projectile proton may experience additional scatterings with other partons from the nucleus before it annihilates into a virtual photon. Taking into account these multiple parton scatterings, one can express the Drell-Yan differential cross section as a sum of contributions from single, double, and higher multiple scatterings. 
In the single scattering as illustrated in Fig.~\ref{fig:general}(a), a single parton with momentum $x'p'$ from the proton scatters with only one parton with momentum $k=xp$ from the nucleus participating in the hard collision. Such a single hard scattering is localized in space and time, and thus does not lead to significant medium size enhanced modification to the production rate from p+p to p+A collisions, other than a mild $A$ dependence from nuclear parton distribution functions.
On the other hand, for the double scattering as illustrated in Fig.~\ref{fig:general}(b), the single parton from the proton could scatter off two partons (with momenta $k_1,~k_g$ in the amplitude) simultaneously in the nucleus. Such a double scattering is usually power suppressed by the hard scale, but it could be enhanced by the nuclear size $\sim A^{1/3}$, which arises when two partons come from different nucleons inside the nucleus. It is such double scattering (as well as higher multiple scattering) that leads to significant difference in $\langle q_T^2\rangle$ between p+A and p+p collisions, and thus the major contribution to the transverse momentum broadening $\Delta\langle q_T^2\rangle$. In this paper, we will focus on the double scattering contribution to the transverse momentum broadening, 
\bea
\Delta \langle q_{T}^2\rangle \approx \left. \frac{d\langle q_{T}^2\sigma^{D}\rangle}{dQ^2}
 \right/ \frac{d\sigma}{dQ^2},
\qquad
{\rm with~~}
\frac{d\langle q_{T}^2\sigma^D\rangle}{dQ^2} \equiv \int dq_{T}^2q_{T}^2 \frac{d\sigma^D}{dQ^2dq_{T}^2}.
 \label{eq-weight}
\eea
Here the superscript ``D'' represents the double scattering contribution. That is, we need to calculate
the transverse momentum weighted differential cross section $d\langle q_{T}^2\sigma^D\rangle/dQ^2$, which is the main focus of our paper.  We will calculate it at both LO and NLO in the expansion of the strong coupling constant $\alpha_s$ in QCD perturbation theory.
\bef
\psfig{file=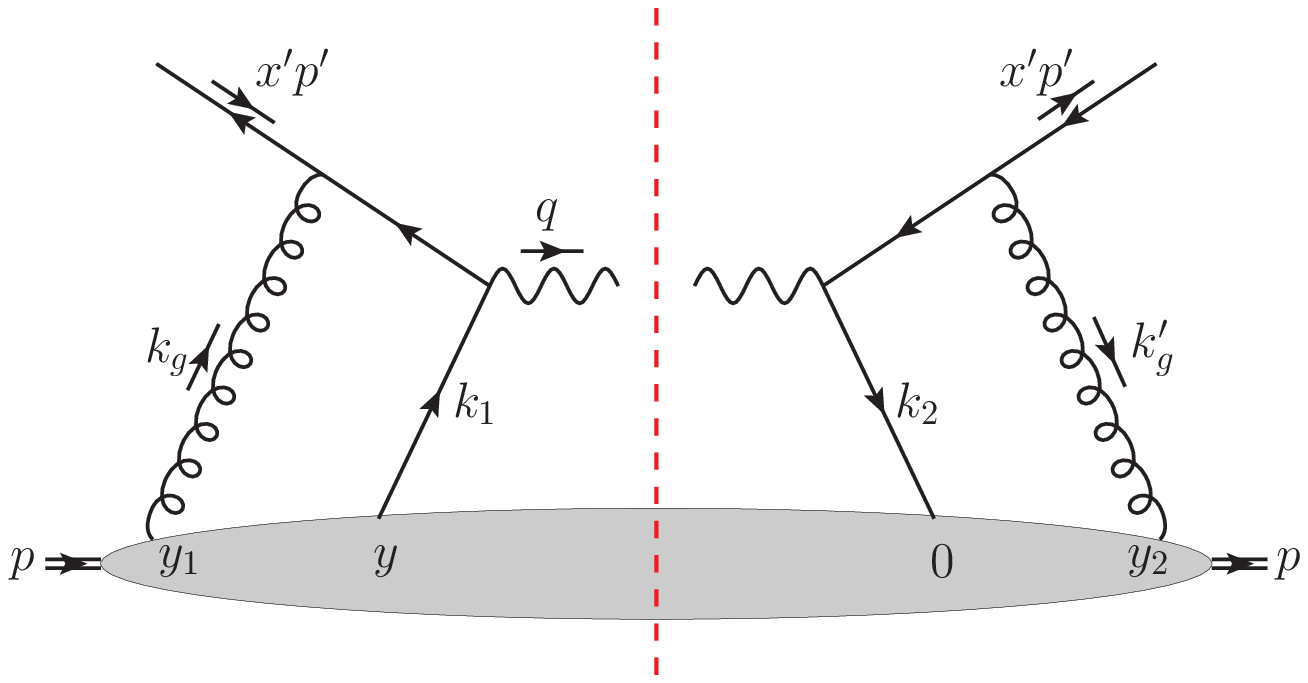, width=3in}
\caption{Feynman diagram for double scattering contribution to transverse momentum broadening at LO. The momenta for the initial quarks and gluons are $k_1=x_1p$, $k_2=(x_1+x_3)p$, $k_g=x_2p+k_T$ and $k_g'=(x_2-x_3)p+k_T$, respectively.}
\label{fig-LO}
\eef

All our calculations are performed in a covariant gauge. The Feynman diagram for the double scattering contribution at LO is shown in Fig.~\ref{fig-LO}, where an anti-quark with momentum $x'p'$ from the proton undergoes an additional scattering in the nucleus by exchanging a gluon with momentum $k_g$ before it annihilates with the quark of momentum $k_1$ into the virtual photon~\footnote{We also consider the case in which the quark from the proton scatters off an anti-quark from the nucleus, and find that the hard parts of this channel are exactly the same as what we have discussed here for both LO and NLO. Thus these contributions are included in our final result by understanding $\sum_q$ runs over both quark and anti-quark flavors.}. The part of the diagram on the right hand side of the vertical dashed line in Fig.~\ref{fig-LO} is the complex conjugate of the amplitude. The LO computation is rather straightforward, and has been explained in details in Refs.~\cite{Guo:1998rd,Fries:2002mu,Kang:2008us,Kang:2011bp,Xing:2012ii,Kang:2014ela}. Here we present only the final result and neglect all the technical details. To be consistent with the calculations at NLO, we present the LO result as calculated in $n=4-2\epsilon$ dimensions:
\bea
\frac{d\langle q_{T}^2\sigma\rangle^{(\rm LO)}}{dQ^2}=\sigma_{\ell}\sum_q\int\frac{dx'}{x'}f_{\bar q/p}(x')\int\frac{dx}{x}T_{gq}(x,0,0)\delta(1-z),
\label{eq-weight-lo}
\eea
where 
\bea
\sigma_{\ell}=\sigma_0\left(\frac{4\pi^2\alpha_s}{N_c}\right),
\qquad
{\rm with~~}
\sigma_0=\frac{4\pi\alpha_{\rm em}^2}{9Q^2}(1-\epsilon),
\eea
and $\alpha_{\rm em}$ is the fine structure constant; and $\sum_q$ runs over all quark flavors, $f_{\bar q/p}(x')$ is the anti-quark distribution function inside the proton, and $T_{gq}(x, 0, 0)$ is a twist-4 gluon-quark correlation function of the colliding heavy ion. In Eq.~\eqref{eq-weight-lo}, $\sigma_\ell$, is the lowest order partonic cross section for DY process; $\delta(1-z)$ with $z=x_B/x$ and the Bjorken variable of DY process, $x_B=Q^2/2p\cdot q$, defines the differential cross section in $Q^2$; and the rest part of the right-hand-side represents the effective partonic flux and the broadening at the level of double scattering, which was encoded in the twist-4 gluon-quark correlation function, 
$T_{gq}$,  which has the following operator definition \cite{Guo:1998rd}
\bea
T_{gq}(x_1, x_2, x_3)
=&\int \frac{dy^-}{2\pi} e^{ix_1p^+y^-}  \int \frac{dy_1^-dy_2^-}{4\pi} e^{ix_2p^+(y_1^- - y_2^-)}
 e^{ix_3p^+y_2^-} \theta(-y_2^-)\theta(y^- - y_1^-)
\nnu
&
\times  \langle A| F_{\sigma}^+(y_2^-) {\bar\psi}_q(0) \gamma^+ \psi_q(y^-) F^{\sigma +}(y_1^-) |A\rangle,
\label{Tqg}
\eea
with the $\theta$-functions taking care of the order of the double scatterings in DY process. With the usual partonic Mandelstam variables for DY process defined as follows:
\bea
\hat s=(x'p'+xp)^2,
\qquad
\hat t=(xp-q)^2,
\qquad
\hat u=(x'p'-q)^2,
\label{mandelstam}
\eea
we have the Lorentz invariant variable $z=x_B/x=Q^2/\hat{s}$. Substituting the LO result of the $q_T^2$-weighted cross section as shown in Eq.~\eqref{eq-weight-lo} into the definition of transverse momentum broadening in Eq.~\eqref{eq-weight}, and performing the $x$ integration by using $\delta(1-z)$, we obtain $\Delta\langle q_T^2\rangle$ at LO:
\bea
\Delta \langle q_{T}^2\rangle = \left(\frac{4\pi^2\alpha_s}{N_c}\right)\frac{\sum_qe_q^2\int\frac{dx'}{x'}f_{\bar q/p}(x')T_{gq}(x_B,0,0)}
{\sum_qe_q^2\int\frac{dx'}{x'}f_{\bar q/p}(x')f_{q/A}(x_B)}\, .
\eea
This result is consistent with earlier calculations in Refs.~\cite{Guo:1998rd,Fries:2002mu,Kang:2008us}.

\section{Transverse momentum broadening at next-to-leading order}
\label{sec-NLO}
In this section, we present our NLO calculations of the transverse momentum broadening in DY process. We first study the double scattering contributions to the antiquark-quark annihilation channel, $\bar q+q\to g+\gamma^*$, which involve gluon-quark correlation function $T_{gq}$ as defined in Eq.~\eqref{Tqg}. We then derive the result for quark-gluon Compton scattering channel, $q+g\to q+\gamma^*$, which involves gluon-gluon correlation function $T_{gg}$ as defined in Eq.~\eqref{Tgg} below. The final result will be presented at the end of this section, where a QCD evolution equation involving $T_{gq}$ and $T_{gg}$ will also be identified. 

To derive the double scattering contribution to the cross section, i.e. $d\langle q_T^2\sigma^D\rangle/dQ^2$, we follow the well-established generalized high-twist factorization formalism~\cite{Luo:1992fz,Luo:1994np,Qiu:1990xy}. Within such a formalism, one has to perform the so-called collinear expansion, i.e., expand the hard scattering contribution around the vanishing parton transverse momentum. For the generic Feynman diagram as shown in Fig.~\ref{fig:general}(b), the double scattering contribution to the Drell-Yan cross section in $n=4-2\epsilon$ dimensions is given by 
\bea
\frac{d\langle q_T^2\sigma^D\rangle}{dQ^2}=&\frac{1}{2s}\frac{2\alpha_{\rm em}}{3Q^2}\sum_q\int\frac{dx'}{x'}f_{\bar q/p}(x')\int\frac{dy^-}{2\pi}\frac{dy_1^-}{2\pi}\frac{dy_2^-}{2\pi}
\frac{1}{2}\langle A|F_{\sigma}^+(y_2^-)\bar{\psi}_q(0)\gamma^+\psi_q(y^-)F^{\sigma +}(y_1^-)|A\rangle 
\nnu
&\times\left[-\frac{1}{2(1-\epsilon)}g^{\alpha\beta}\right]
\left[\frac{\partial^2}{\partial k_{2T}^{\alpha}\partial k_{3T}^{\beta}}
{\overline H}_{\mu\nu}(k_{2T},k_{3T},p',p,q,\{y_i\})\right]_{k_{2T}=k_{3T}=0}(-g^{\mu\nu})\, ,
\label{eq-general}
\eea
where $\{y_i\}=\{y^-,y^-_1,y^-_2\}$.  In Eq.~(\ref{eq-general}), the factor $2\alpha_{\rm em}/3Q^2$ comes from the leptonic decay $\gamma^*\to \ell^+\ell^-$~\cite{Kang:2012vm,Kang:2008wv,Berger:2001wr}, $(-g^{\mu\nu})$ is the polarization sum for the virtual photon, and 
${\overline H}_{\mu\nu}(k_{2T},k_{3T},,p',p,q,\{y_i\})$ is the Fourier transform of the hard partonic function 
$H_{\mu\nu}(\{x_i\},k_{2T},k_{3T},p',p,q)$,
\bea
{\overline H}_{\mu\nu}(k_{2T},k_{3T},p',p,q,\{y_i\})=\int dx_1dx_2dx_3e^{ix_1p^+y^-}e^{ix_2p^+(y_1^- - y_2^-)}e^{ix_3p^+y_2^-}
H_{\mu\nu}(\{x_i\},k_{2T},k_{3T},p',p,q)\, d{\rm PS}^{(2)},
\eea
where $\{x_i\}=\{x_1,x_2,x_3\}$ are the independent longitudinal momentum fractions carried by the partons from nucleus, $k_{2T},~k_{3T}$ are the two independent transverse momenta for the initial two gluons from nucleus, and $d{\rm PS}^{(2)}$ is the final-state photon-gluon two-particle phase space, which will be defined later in Eq.~(\ref{eq-PS}). To preserve momentum conservation, one has to assign $k_{2T}-k_{3T}$ to one of the initial quarks. We will use gauge invariance to guide the transverse momentum flow for these expansion momenta $k_{2T}$ and $k_{3T}$, which are explained in details below for each different double scattering subprocesses.

\subsection{gluon-quark double scattering} 
In gluon-quark double scattering, before the hard interaction of quark-antiquark annihilation, the initial antiquark from the projectile proton undergoes additional scattering with the nuclear medium through the exchange of a gluon, as illustrated in Fig.~\ref{fig:general}(b). According to the momentum of this exchanged gluon (i.e., $k_g$ and $k_g'$ on the amplitude and complex conjugate of the amplitude in Fig.~\ref{fig:general}(b), respectively), we classify the gluon-quark double scattering into four subprocesses: soft-soft, hard-hard, soft-hard and hard-soft double scatterings. For example, soft-hard refers to the situation where the exchanged gluon has {\it zero} and {\it finite} momenta in the amplitude and the complex conjugate, respectively. In other words, $k_g= 0$ and $k_g'\neq 0$ when taking the $k_{2T}=k_{3T}=0$ limit in Fig.~\ref{fig:general}(b). Below we discuss how to compute these four subprocesses. We first study the central-cut diagrams where exactly one exchanged gluon is on each side of the cut, and we then perform the calculations for the single-triple scattering interference diagrams where two exchanged gluons are on the same side of the cut line. 

\subsubsection{soft-soft double scattering}
\label{ss-gq}
The central-cut diagrams for soft-soft gluon-quark double scattering contributions are shown in Fig.~\ref{fig-gq-ss}, where the ``H''-blobs present the hard $2\to 2$ process $\bar q+q\to g+\gamma^*$ as shown in Fig.~\ref{fig-qq2gphoton}. The short bars indicate the propagators where the soft poles arise, which put the associated quarks on their mass-shell. One thus might consider such a double scattering as two ``factorized'' scatterings: the first $\bar q+g \to \bar q$ through a soft gluon exchange, followed by $\bar q + q \to g+\gamma^*$. To ensure this second $\bar q+q\to g+\gamma^*$ scattering gauge invariant, the quark from the nucleus has to be on mass-shell, i.e., $k_1^2=k_2^2=0$~\cite{Peskin:1995ev}. The most convenient way in performing twist-4 calculation of soft-soft double scattering is to set $k_g=x_2p+k_T$ and $k_g'=(x_2-x_3)p+k_T$, following the set-up of the original high-twist expansion as developed by Qiu and Sterman~\cite{Luo:1994np}. In other words, $k_{2T}=k_{3T}=k_T$, therefore no transverse momentum flows into/out from the initial quark. In this case, the two initial quarks from nucleus have only $``+"$ component with $k_1=x_1p$ and $k_2=(x_1+x_3)p$, thus they are on their mass shell and satisfy the requirements for the gauge invariance. To avoid double counting in symmetrizing the transverse momentum flow, an additional factor $1/2$ is needed in the collinear expansion:
\bea
\left[\frac{\partial^2}{\partial k_{2T}^{\alpha}\partial k_{3T}^{\beta}}
{\overline H}_{\mu\nu}(k_{2T},k_{3T},p',p,q,\{y_i\})\right]_{k_{2T}=k_{3T}=0}
\to
\left[\frac{1}{2}\frac{\partial^2}{\partial k_{T}^{\alpha}\partial k_{T}^{\beta}}
{\overline H}_{\mu\nu}(k_{T},p',p,q,\{y_i\})\right]_{k_{T}=0}\,.
\eea

\bef
\psfig{file=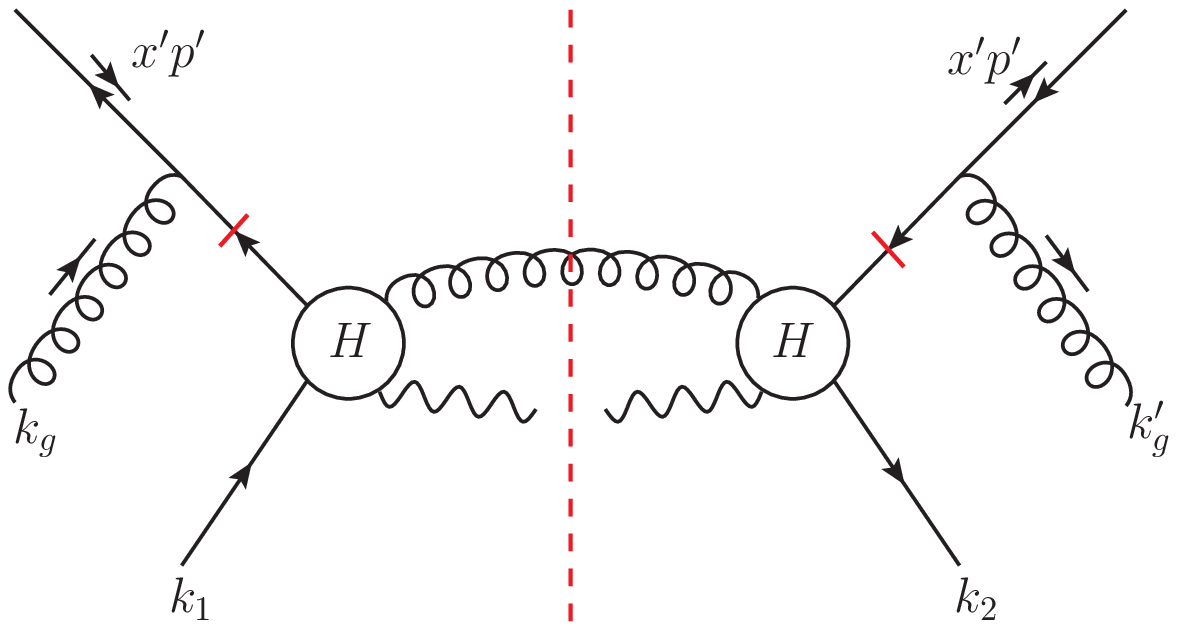, width=3in}
\caption{The central-cut diagrams for soft-soft gluon-quark double scattering in DY process. The short bars indicate the propagators where the soft poles arise. The ``H"-blobs represent the hard $2\to 2$ processes $\bar q+q \to g + \gamma^*$ as shown in Fig. \ref{fig-qq2gphoton}. The parton momenta for the four initial partons are $k_1=x_1p$, $k_2=(x_1+x_3)p$, $k_g=x_2p+k_T$ and $k_g'=(x_2-x_3)p+k_T$.}
\label{fig-gq-ss}
\eef

\bef
\psfig{file=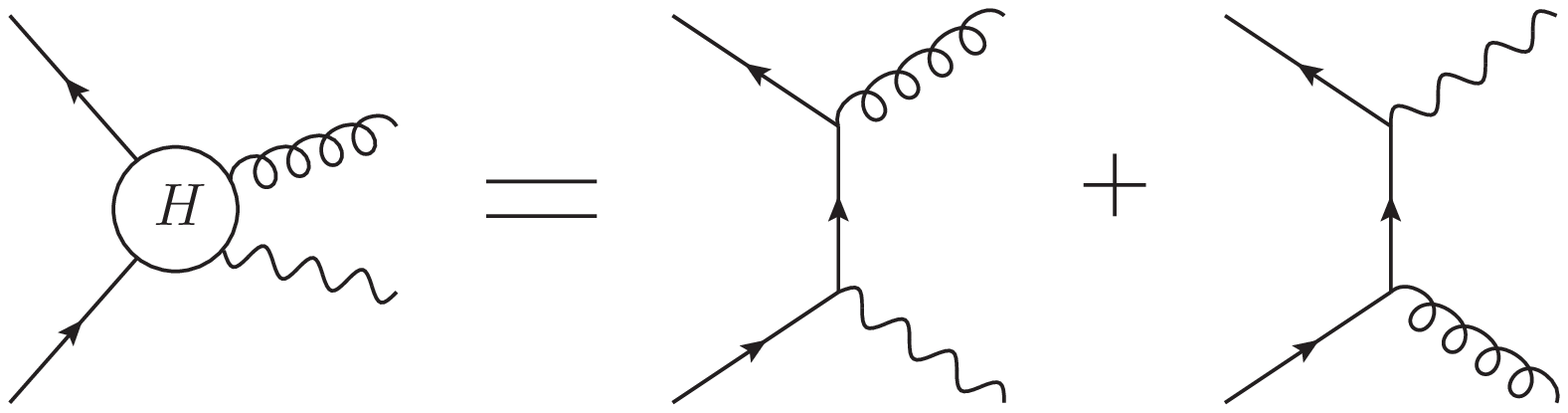, width=3.5in}
\caption{The representations of hard $2\to 2$ processes for $\bar q+q  \to \gamma^*+g$.}
\label{fig-qq2gphoton}
\eef

In soft-soft double scattering, the soft-poles arise from the propagators marked by the short bars in Fig.~\ref{fig-gq-ss}, which can be expressed in the following forms:
\bea
&\frac{1}{(x'p'+k_g)^2+i\epsilon} = \frac{x}{{\hat s}}\frac{1}{x_2+x_c+i\epsilon},\nnu
&\frac{1}{(x'p'+k_g')^2-i\epsilon} = \frac{x}{{\hat s}}\frac{1}{x_2-x_3+x_c-i\epsilon},
\label{eq-pole}
\eea
with $x_c=x\,k_T^2/{\hat s}$. Performing contour integrations, these two propagators can be used to fix the momentum fractions, $x_2$ and $x_3$. At the same time, $x_1$ is fixed by the on-shell condition of the unobserved final-state gluon, which is characterized by the $\delta$-function in the two-body final-state phase space,
\bea
d{\rm PS}^{(2)}&=\frac{1}{2\pi}\int dx\delta\left(x_1+x_2-x-x_e\right)\int\frac{d^nq}{(2\pi)^{n-2}}\delta(q^2-Q^2)\delta[(x'p'+xp-q)^2],
\nnu
&=\frac{1}{16\pi^2}\int dx\delta\left(x_1+x_2-x-x_e\right)
\left(\frac{4\pi}{Q^2}\right)^{\epsilon}
\frac{1}{\Gamma(1-\epsilon)}z^{\epsilon}(1-z)^{1-2\epsilon}\int_0^1dv[v(1-v)]^{-\epsilon}.
\label{eq-PS}
\eea
Here $x_e=x\,(k_T^2-2q\cdot k_T)/{\hat u}$, and $v=(1+\cos\theta)/2$ with $\theta$ the partonic center-of-mass angle. 
By performing contour integrations of the un-pinched soft poles as in Eq.~\eqref{eq-pole}, and combining with the final state phase space as in Eq.~\eqref{eq-PS}, we then fix all the parton momentum fractions as 
\bea
x_1=x+x_e+x_c,
\qquad
x_2=-x_c,
\qquad
x_3=0\, ,
\label{eq-fix-x}
\eea
with $x_c=x\,k_T^2/{\hat s}$ defined earlier.
One can immediately see that in the collinear limit $k_T\to 0$, the momenta for the two exchanged gluons are zero: $k_g\to 0,~k_g'\to 0$ due to $x_2\to 0$, $x_2-x_3\to 0$ in this limit. This is why we refer to this process as soft-soft double scattering. 

As shown in Eq. (\ref{eq-general}), the key step in the high-twist factorization formalism is to perform the collinear expansion, from which we obtain the gauge invariant result for physical double scatterings.
In general, the collinear expansion involves not only the hard part, but also the twist-4 matrix element because of the $k_T$-dependence of momentum fractions $x_i$ as shown in Eq.~\eqref{eq-fix-x}. Taking both into account, we obtain the following generic expression,
\bea
\frac{\partial^2 [T_{gq}(\{x_i\})H(\{x_i\},k_T)]}{\partial k_T^{\alpha}\partial k_T^{\beta}}
=&\Bigg\{x^2\frac{d^2 T_{gq}(x,0,0)}{d x^2}\left[\left(\frac{2}{\hat u}\right)^2\frac{q_T^2}{n-2}H\right]
+x\frac{d T_{gq}(x,0,0)}{d x}\left[2\left(\frac{1}{\hat u}+\frac{1}{\hat s}\right)H-\frac{4}{\hat u}\frac{q_T^2}{n-2}\frac{\partial H}{\partial w_1}\right]
\nnu
&
+T_{gq}(x,0,0)\left[\frac{q_T^2}{n-2}\frac{\partial^2H}{\partial w_1^2}+2\frac{\partial H}{\partial w_2}\right]\Bigg\}\, g_T^{\alpha\beta},
\label{eq-exp-ss}
\eea
where $n=4-2\epsilon$ is the dimension of the space-time, $w_1=q\cdot k_T$, $w_2=k_T^2$ and the arguments in the hard function $H$ are suppressed.

The next step is to perform the $\epsilon$-expansion and identify the divergences. Realizing the partonic Mandelstam variables defined in Eq.~\eqref{mandelstam} can be written as
\bea
\hat s = \frac{Q^2}{z}, 
\qquad
\hat t = -\frac{Q^2(1-z)(1-v)}{z},
\qquad
\hat u = -\frac{Q^2 (1-z) v}{z},
\eea
as well as $q_T^2 = \hat t\hat u/\hat s$, from the final-state phase space in Eq.~\eqref{eq-PS}, one simply has to pay extra attention to the expansion around $z\to 1$, $v\to  1$ and $v\to 0$, where divergences could be encountered. It is instructive to realize that $z\to 1$ corresponds to the phase space where the radiated gluon is soft and leads to a soft divergence. Such a soft divergence cancels out between real and virtual diagrams as we will demonstrate below. On the other hand, $v=(1+\cos\theta)/2 \to1$ corresponds to $\theta \to 0$, i.e., the radiated gluon is collinear to the projectile proton; while $v\to 0$ (or $\theta\to \pi$) stands for the situation where the radiated gluon is collinear to the colliding nucleus. That is, both limits of $v$ lead to perturbative collinear divergences. For later convenience, let us present the weighted cross section in the following generic form:
\bea
\frac{d\langle q_T^2\sigma^D\rangle^{ij}}{dQ^2} &= \sigma_{\ell} \frac{\alpha_s}{2\pi}\sum_q
\int\frac{dx'}{x'}f_{\bar q/p}(x')
\int_{x_B}^{1}\frac{dx}{x}\int_0^1dv\left(\frac{4\pi\mu^2}{Q^2}\right)^{\epsilon}
\frac{1}{\Gamma(1-\epsilon)}\left\{\frac{1}{\epsilon^2}\delta(1-z)\left[\delta(v)+\delta(1-v)\right]T_{gq}\otimes H_{d2}^{ij}
\right.\nnu
&\left.-\frac{1}{\epsilon}\delta(1-v)T_{gq}\otimes H_{d1}^{ij}
-\frac{1}{\epsilon}\delta(v)T_{gq}\otimes \tilde H_{d1}^{ij}+ T_{gq}\otimes H_{gq-C}^{ij}
\right\},
\label{eq:master}
\eea
where the superscript ``$i$'' and ``$j$'' represent ``s'' and ``h'', indicating contributions from soft and hard rescatterings. Notice that one can easily perform the integral over $v$ for the divergent terms, but here we still keep it unintegrated, so that we can keep track of different phase spaces for the radiated gluon, and regularize them accordingly in the factorization step. For the derivative terms (the first and second terms in Eq.~\eqref{eq-exp-ss}), we further perform integration by parts to convert them into the form of non-derivative terms, thus we can combine them with the rest of the non-derivative terms properly. Following the techniques as outlined in Ref.~\cite{Kang:2014ela}, we arrive at the following divergent terms for soft-soft double scattering, 
\bea
T_{gq}\otimes H_{d2}^{ss}=T_{gq}(x,0,0)C_F, 
\qquad
T_{gq}\otimes H_{d1}^{ss}=T_{gq}(x,0,0)C_F\frac{z^2(1+z^2)}{(1-z)_+},
\qquad
T_{gq}\otimes \tilde H_{d1}^{ss}=T_{gq}(x,0,0)C_F\frac{1+z^2}{(1-z)_+},
\eea
where the superscript ``ss'' indicates soft-soft double scattering contributions, and the ``plus''-function is defined as 
\bea
\int_0^1 dz\frac{f(z)}{(1-z)_+} \equiv \int_0^1 dz\frac{f(z)-f(1)}{1-z}. 
\eea
On the other hand,  the finite contribution at the NLO denoted by $T_{gq}\otimes H_{gq-C}^{ss}$ is collected in the Appendix, as given by Eq.~\eqref{eq-gq-ss}.

\subsubsection{hard-hard double scattering}
Let us now study the contributions from the hard-hard double scatterings, with the central-cut diagrams given in Fig.~\ref{fig-gq-hh} where the ``H"-blobs represent the hard $2\to 2$ processes $\bar q+g \to g+\bar q$ as shown in Fig.~\ref{fig-qg2qg}. The crosses in Fig.~\ref{fig-gq-hh} indicate the propagators where the hard poles arise, which put the associated quarks on their mass-shell. Thus one might consider that the radiated gluon is induced by the first antiquark-gluon hard interaction, $\bar q + g\to g+\bar q$, followed by the annihilation process $\bar q+q\to \gamma^*$ to generate the virtual photon. 
In this case, to ensure the gauge invariance for the physical double scattering, in particular the first $\bar q + g\to g+\bar q$ process, the exchanged gluons have to be on their mass-shell: at least up to the order of our collinear expansion. In other words, one should have $k_g^2=0$ up to $\mathcal O(k_{2T})$ and $k_g'^2=0$ up to $\mathcal O(k_{3T})$. Thus we can neglect the ``minus" components of the momenta for both exchanged gluons, since they contribute to $\mathcal O(k_{2T}^2)$ and $\mathcal O(k_{3T}^2)$. In other word, we could simply set the momenta of these two exchanged gluons as $k_g=x_2p+k_{2T}$ and $k_g'=(x_2-x_3)p+k_{3T}$. At the same time, to preserve the momentum conservation on the two sides of the cut, we have to assign a transverse momentum of $k_{2T}-k_{3T}$ to the quark from the nucleus either on the left or right side of the cut. We have checked that they lead to exactly the same result~\footnote{Note that these assignments are different from the previous setup as given in Refs.~\cite{Fries:2002mu,Xing:2011fb,Guo:1997it}, in which one simply sets $k_{2T}=k_{3T}=k_T$, i.e. the same as that in soft-soft double scattering given above. We have checked that our new assignments give the same results as the previous setup.}.

\bef
\psfig{file=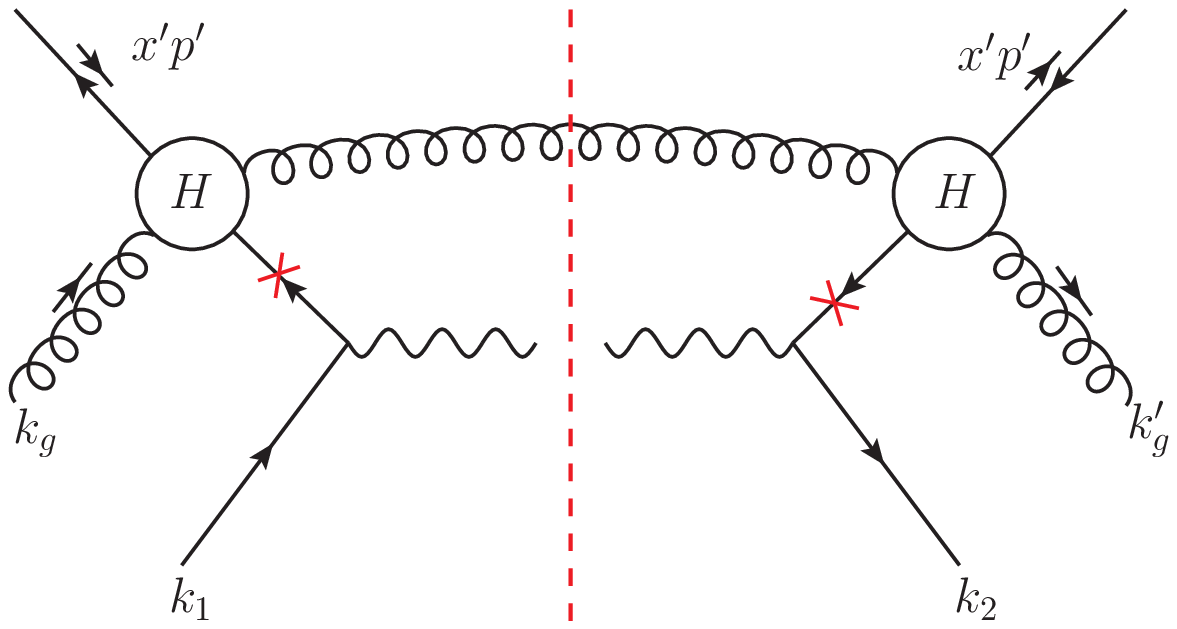, width=3.0in}
\caption{The central-cut diagrams for hard-hard gluon-quark double scattering in DY process. The crosses indicate the propagators where the hard poles arise. The ``H"-blobs represent the hard $2\to 2$ processes $\bar q+g \to g +  \bar q$ as shown in Fig. \ref{fig-qg2qg}. The parton momenta for the four initial partons are $k_1=x_1p$, $k_2=(x_1+x_3)p+k_{2T}-k_{3T}$, $k_g=x_2p+k_{2T}$ and $k_g'=(x_2-x_3)p+k_{3T}$, respectively.}
\label{fig-gq-hh}
\eef

\bef
\psfig{file=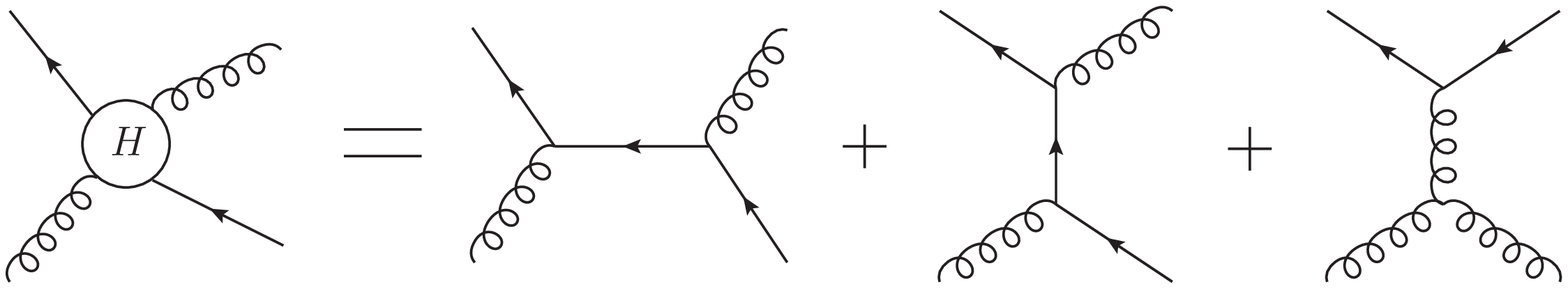, width=4.9in}
\caption{The representations of hard $2\to 2$ processes for $\bar q+g\to g+\bar q$.}
\label{fig-qg2qg}
\eef

In the following, we choose $k_1=x_1p$ and $k_2=(x_1+x_3)p+k_{2T}-k_{3T}$ to demonstrate our computations for the hard-hard double scattering. Similar to the soft-soft double scattering case, the parton momentum fractions $x_{1,2,3}$ can be fixed by the hard poles as well as the on-shell condition of the unobserved parton (i.e. the radiated gluon):
\bea
x_1=x_d,
\qquad
x_2=x-x_d+x_h,
\qquad
x_3=x_f+x_g,
\eea
where $x_d,~x_f,~x_g,~x_h$ are given by 
\bea
x_d=x\frac{Q^2}{Q^2-\hat t}
\qquad
x_f=x\frac{2q\cdot k_{2T}-2q\cdot k_{3T}}{\hat t-Q^2},
\qquad
x_g =x\frac{2k_{2T}\cdot k_{3T}}{\hat t- Q^2},
\qquad
x_h=x\frac{-2q\cdot k_{2T}}{\hat u}. 
\label{eq:xi-def}
\eea
In the collinear limit $k_{2T}=k_{3T}=0$, we have
\bea
x_1=x\frac{z}{1-v(1-z)},
\qquad
x_2=x\frac{(1-v)(1-z)}{1-v(1-z)},
\qquad
x_3=0,
\label{eq:h-x}
\eea
and thus the momentum fractions of the exchanged gluons ($x_2$ and $x_2-x_3$) are both finite. This is why we refer to this process as hard-hard double scattering. 

Performing the same collinear expansion, we have the following generic result for the hard-hard double scattering:
\bea
\frac{\partial^2 [T_{gq}(\{x_i\})H(\{x_i\},k_T)]}{\partial k_{2T}^{\alpha}\partial k_{3T}^{\beta}}
=T_{gq}(x_1,x_2,x_3)\left[\frac{q_T^2}{n-2}\frac{\partial^2H}{\partial w_3 \partial w_4}+\frac{\partial H}{\partial w_5}\right]g_T^{\alpha\beta},
\label{eq-expn-hh}
\eea
where the parton momentum fractions $x_{1,2,3}$ on the right hand side are those in Eq.~\eqref{eq:h-x}, and the variables $w_3,~w_4,~w_5$ are given by
\bea
w_3=q\cdot k_{2T}, 
\qquad
w_4=q\cdot k_{3T},
\qquad
w_5=k_{2T}\cdot k_{3T}.
\label{eq:yi-def}
\eea
Notice that we do not have derivative terms in hard-hard double scattering compared with the soft-soft double scattering in Eq.~\eqref{eq-exp-ss}, because of the following facts:
\bea
\left.\frac{\partial H}{\partial k_{2T}^{\alpha}}\right|_{k_{2T}=0} = \left.\frac{\partial H}{\partial k_{3T}^{\alpha}}\right|_{k_{3T}=0} = \left.H\right|_{k_{2T/3T}=0}=0\, ,
\eea
which are the results of gauge invariance of the hard scattering amplitude, $\bar{q}+g\to g + \bar{q}$, in Fig.~\ref{fig-qg2qg}, and the fact that the Lorentz indices of the double scattering gluons of momentum $k_g$ and $k'_g$ are contracted by the averaged nuclear momentum vector $p$ that is collinear to the $k_g$ and $k'_g$ when $k_{2T}$ and $k_{3T}\to 0$. 
Using the same notation as in Eq.~\eqref{eq:master}, we present the divergent terms when $\epsilon\to 0$ for hard-hard double scattering as
\bea
T_{gq}\otimes H_{d2}^{hh} & =T_{gq}(x,0,0)C_A,
\qquad
T_{gq}\otimes H_{d1}^{hh}  =T_{gq}(x,0,0)\frac{1+z^2}{(1-z)_+}\left[C_Az+C_F(1-z)^2\right],
\nnu
T_{gq}\otimes \tilde H_{d1}^{hh} &=T_{gq}(xz,x(1-z),0)C_A\frac{2}{(1-z)_+},
\label{eq:hh-div}
\eea
where again the superscript ``hh'' represents the hard-hard double scattering contributions. On the other hand, the finite term for hard-hard double scattering denoted by $T_{gq}\otimes H_{gq-C}^{hh}$ is expressed explicitly in Eq.~\eqref{eq-gq-hh} in Appendix. It might be worthwhile mentioning that when $v\to 1$ or $z\to 1$, the gluon-quark correlation function $T_{gq}(x_1,x_2,x_3)\to T_{gq}(x, 0, 0)$ in Eq.~\eqref{eq-expn-hh}, and we can thus utilize this feature to combine the divergent terms associated with $\delta(1-v)$ and $\delta(1-z)$ in hard-hard and soft-soft double scattering contributions. 

\subsubsection{soft-hard and hard-soft double scattering}
We will now discuss the interference diagrams between the soft and hard rescatterings. The relevant central-cut diagrams are shown in Fig.~\ref{fig-gq-int}, with soft-hard (left) and hard-soft (right) gluon-quark double scatterings, respectively. We explain in details how to calculate the soft-hard double scattering contributions, and only give the results for hard-soft double scattering contribution at the end of this section as the calculation is similar. 

\bef
\psfig{file=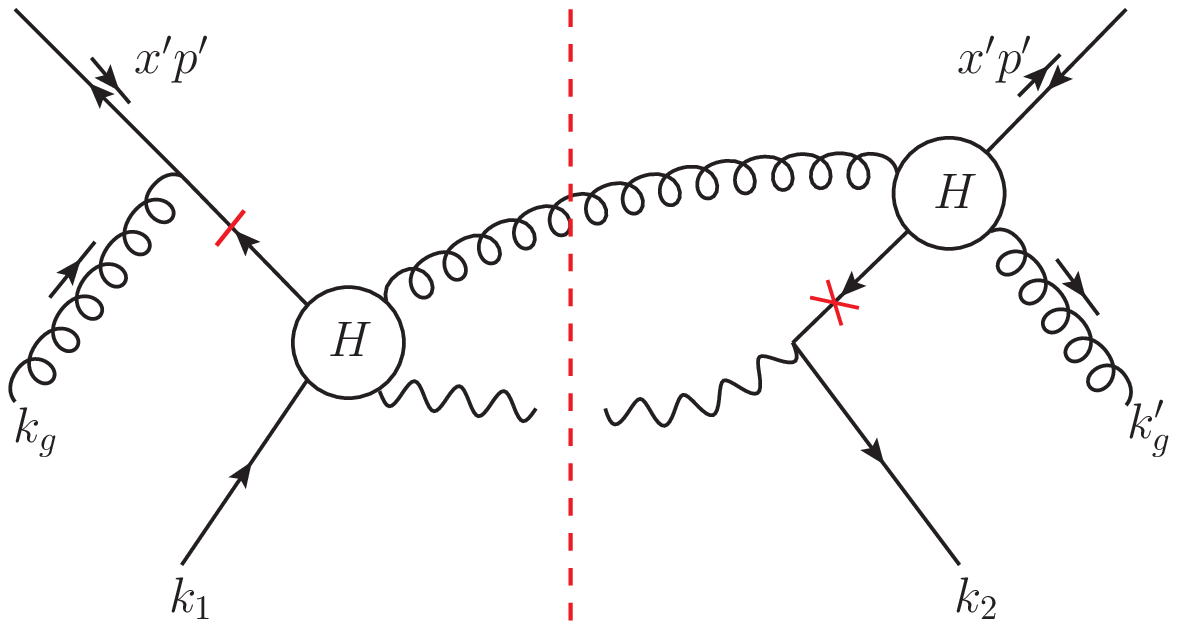, width=3.0in}
\hskip 0.5in 
\psfig{file=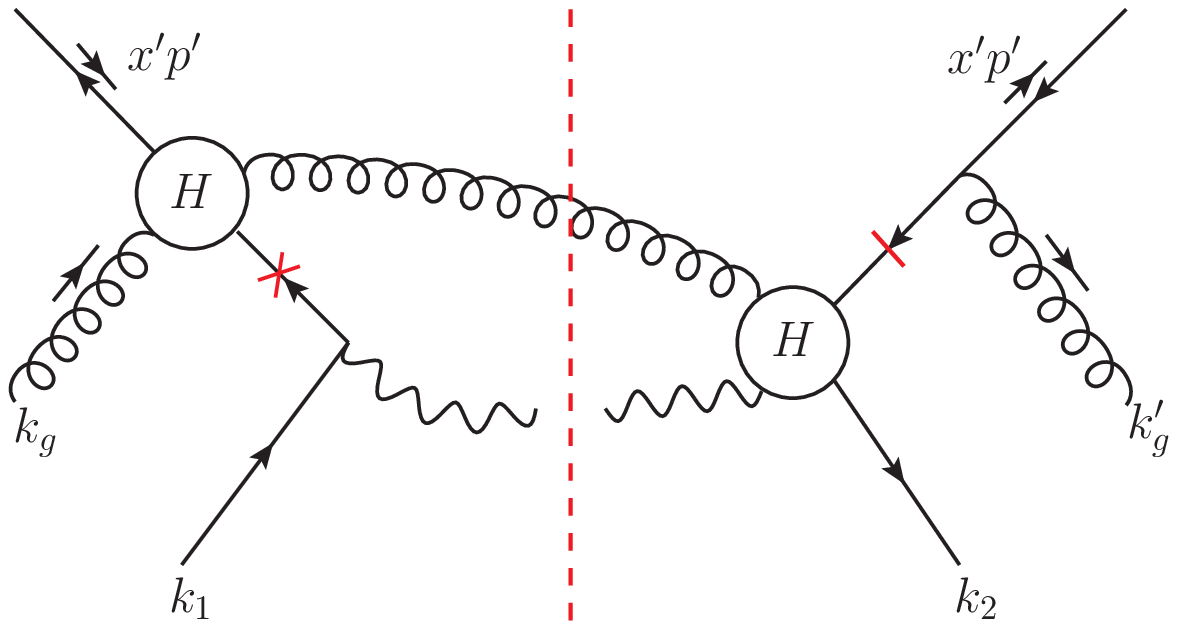, width=3.0in}
\caption{The central-cut diagrams for soft-hard (left) and hard-soft (right) gluon-quark double scattering in DY process. The ``H"-blobs represent the hard $2\to 2$ processes $\bar q+q \to \gamma^* + g$ and $\bar q+g \to \bar q + g$ as shown in Figs. \ref{fig-qq2gphoton} and \ref{fig-qg2qg}, respectively. The parton momenta for the four initial partons are $k_1=x_1p$, $k_2=(x_1+x_3)p+k_{2T}-k_{3T}$, $k_g=x_2p+k_{2T}$ and $k_g'=(x_2-x_3)p+k_{3T}$ for soft-hard double scattering, while $k_1=x_1p+k_{3T}-k_{2T}$, $k_2=(x_1+x_3)p$, $k_g=x_2p+k_{2T}$ and $k_g'=(x_2-x_3)p+k_{3T}$ for hard-soft double scattering.}
\label{fig-gq-int}
\eef

For the soft-hard double scattering as shown in Fig.~\ref{fig-gq-int} (left), the short bar on the left side of the cut represents the propagator where a soft pole arises, while the cross on the right side of the cut stands for the propagator where a hard pole arises. On the left side of the cut (the amplitude), we have first $\bar q+g\to \bar q$ through a soft gluon exchange, followed by $\bar q+q\to g+\gamma^*$. To ensure the gauge invariance, as we have discussed for soft-soft double scattering above, we have to make sure the initial quark from the nucleus to be on mass-shell, i.e. $k_1^2=0$ up to the order of our collinear expansion. At the same time, on the right side of the cut (the complex conjugate of the amplitude), one has a gauge invariant $\bar q+g\to g+\bar q$ partonic scattering subprocess through a hard gluon exchange, followed by $\bar q+q\to \gamma^*$ annihilation. To ensure gauge invariance, as we have discussed for hard-hard double scattering above, we have to make sure that the exchanged gluon to be on mass-shell, i.e. $k_g'^2=0$ up to the order of our collinear expansion. With both requirements, we can assign 
\bea
k_1=x_1p, 
\qquad
k_2=(x_1+x_3)p+k_{2T}-k_{3T}, 
\qquad
k_g=x_2p+k_{2T}, 
\qquad
k_g'=(x_2-x_3)p+k_{3T}.
\eea
With such momentum assignments, we can use the pole propagators and the on-shell condition of the final-state radiated gluon, to fix the parton momentum fractions:
\bea
x_1 = x+x_h,
\qquad
x_2=0,
\qquad
x_3=x_d-x+x_f+x_g-x_h ,
\eea
where $x_h$, $x_d$, $x_f$, and $x_g$ are defined in Eq.~(\ref{eq:xi-def}).
Therefore, the collinear expansion can be expressed as follows,
\bea
\frac{\partial^2 [T_{gq}(\{x_i\})H(\{x_i\},k_T)]}{\partial k_{2T}^{\alpha}\partial k_{3T}^{\beta}}
=&\left\{x\frac{d T_{gq}(x,0,x_d-x)}{d x}\left[\frac{2}{-\hat u}\frac{q_T^2}{n-2}\frac{\partial H}{\partial w_4}\right]
+ x\frac{d T_{gq}(x,0,x_d-x)}{d x_d}\left[\frac{2}{\hat t-Q^2}\frac{q_T^2}{n-2}\frac{\partial H}{\partial w_4}\right]
\right.
\nnu
&\left.+T_{gq}(x,0,x_d-x)\left[\frac{q_T^2}{n-2}\frac{\partial^2H}{\partial w_3 \partial w_4}+\frac{\partial H}{\partial w_5}\right]\right\}g_T^{\alpha\beta},
\label{eq-exp-sh}
\eea
where $w_i$ with $i=3,4,5$ are defined in Eq.~(\ref{eq:yi-def}).
Unlike the collinear expansion in Eq.~(\ref{eq-exp-ss}) for the case of soft-soft double scattering, the expansion for the soft-hard double scattering in Eq.~(\ref{eq-exp-sh}) does not have the term proportional to $d^2 T_{gq}/dx^2$, due to the fact that $H|_{k_{2T/3T}=0}=0$, an immediate consequence of the gauge invariance of the partonic hard scattering part of the soft-hard double scattering.  For the same reason, unlike the expansion for the hard-hard double scattering in Eq.~(\ref{eq-expn-hh}), the collinear expansion for the soft-hard double scattering in Eq.~(\ref{eq-exp-sh}) does have the terms proportional to the linear derivative of the gluon-quark correlation function.
From this collinear expansion formula in Eq.~(\ref{eq-exp-sh}) it is straightforward to derive the divergent terms when $\epsilon \to 0$ for the soft-hard double scattering contribution. Again using the notation in Eq.~\eqref{eq:master}, we have
\bea
&T_{gq}\otimes H_{d2}^{sh}=-T_{gq}(x,0,0)\frac{C_A}{2},
\qquad
T_{gq}\otimes H_{d1}^{sh}=T_{gq}(x,0,0)\frac{z(1+z^2)}{(1-z)_+}\left[-\frac{C_A}{2}+C_F(1-z)\right],
\nnu
&T_{gq}\otimes \tilde H_{d1}^{sh}=-T_{gq}\left(x,0,-x(1-z)\right)\frac{C_A}{2}\left[\frac{1+z}{(1-z)_+} - 2\delta(1-z)\right].
\eea
On the other hand, the associated finite term as denoted by $T_{gq}\otimes H_{gq-C}^{sh}$ is given explicitly in Eq.~\eqref{eq-gq-sh} in Appendix. 

Similarly we compute the hard-soft double scattering contributions. It turns out that the divergent terms associated with $\delta(1-z)$ and $\delta(1-v)$ are exactly the same as those for soft-hard double scattering, while the term associated with $\delta(v)$ can be obtained by the following replacement in the corresponding term of soft-hard double scattering,
\bea
T_{gq}\left(x,0,-x(1-z)\right) \to T_{gq}\left(xz,x(1-z),x(1-z)\right).
\eea
The finite term denoted as $T_{gq}\otimes H_{gq-C}^{hs}$ is given in Eq.~\eqref{eq-gq-hs} in Appendix. 

\subsubsection{virtual corrections}

In this subsection, we calculate the virtual corrections in gluon-quark double scattering, which have to be included to ensure unitarity and infrared safety of the partonic hard parts of the factorized gluon-quark double scattering contribution to the DY transverse momentum broadening. The relevant Feynman diagrams for the virtual corrections are shown in Fig.~\ref{fig-v}, with the black blob given by Fig.~\ref{fig-v-explicit}. Two diagrams in Fig.~\ref{fig-v} are actually the complex conjugate of each other, and they have the same results. The actual calculations of the virtual corrections contain significant amount of tensor reductions and integrations, and are thus quite involved and tedious. Nevertheless, the techniques are quite clear and straightforward as demonstrated for SIDIS process in our previous paper \cite{Kang:2014ela}. The final result is given by
\bea
\frac{d\langle q_T^2\sigma^D\rangle^{\rm (V)}}{dQ^2} &= \sigma_{\ell} \frac{\alpha_s}{2\pi}\sum_q
\int\frac{dx'}{x'}f_{\bar q/p}(x')
\int_{x_B}^{1}\frac{dx}{x}\left(\frac{4\pi\mu^2}{Q^2}\right)^{\epsilon}
\frac{1}{\Gamma(1-\epsilon)}T_{gq}(x,0,0)C_F\delta(1-z)\left[-\frac{2}{\epsilon^2}
-\frac{3}{\epsilon}-8+\pi^2\right],
\label{eq:virtual}
\eea
which differs only by the factor of $\pi^2$ from SIDIS, due to the fact that $Q^2 < 0$ ($Q^2 > 0$) for SIDIS (DY). In other words, it has exactly the same structure as the virtual correction at leading-twist. Such type of behavior also holds true at twist-3 level for transverse momentum weighted spin dependent cross sections as shown in~\cite{Vogelsang:2009pj,Kang:2012ns,VITEV:2014wza}. For later convenience, we will define a finite term for the virtual correction, with the following expression:
\bea
T_{gq}\otimes H^{V}_{gq-C}=C_F\delta(1-z)\left(-8+\pi^2\right)T_{gq}(x,0,0).
\label{eq:virtual-finite}
\eea

\bef
\psfig{file=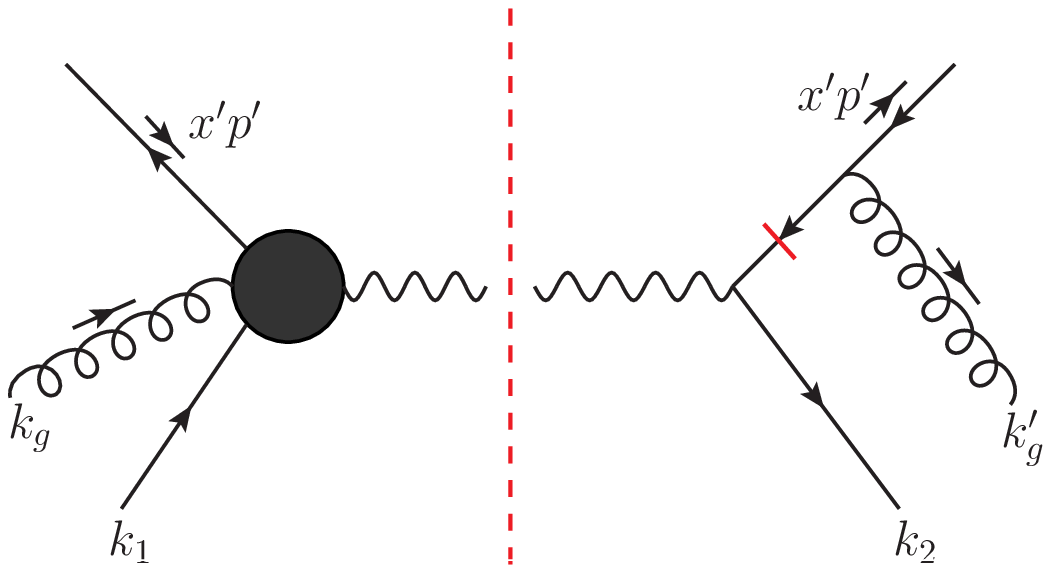, width=2.6in}
\hskip 0.3in 
\psfig{file=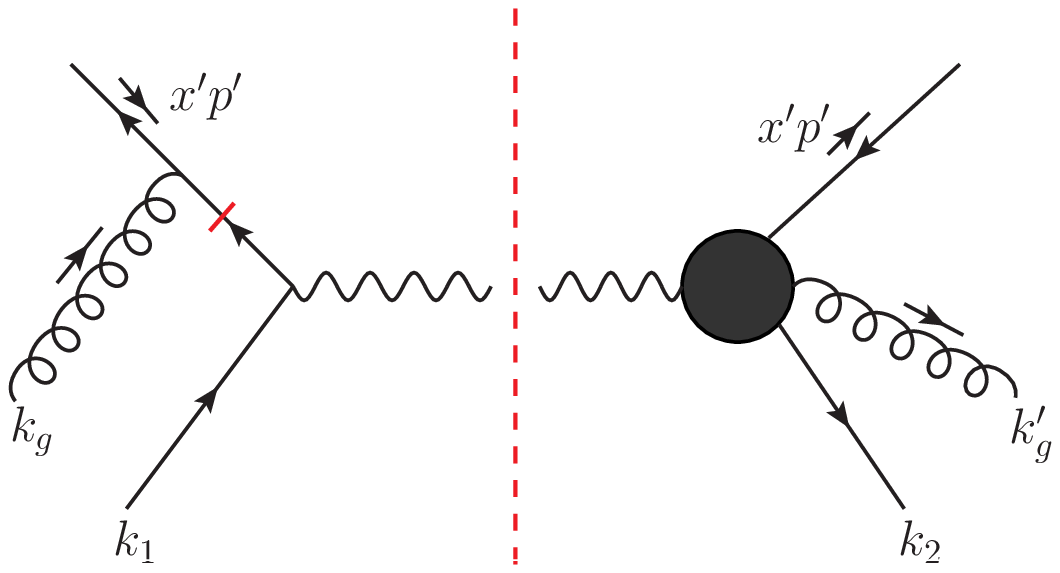, width=2.6in}
\caption{The virtual diagrams in DY at twist-4, the parton momenta follow the same convention as in Fig. \ref{fig-LO}, the blob contains 7 diagrams as in Fig. \ref{fig-v-explicit}. }
\label{fig-v}
\eef

\bef
\psfig{file=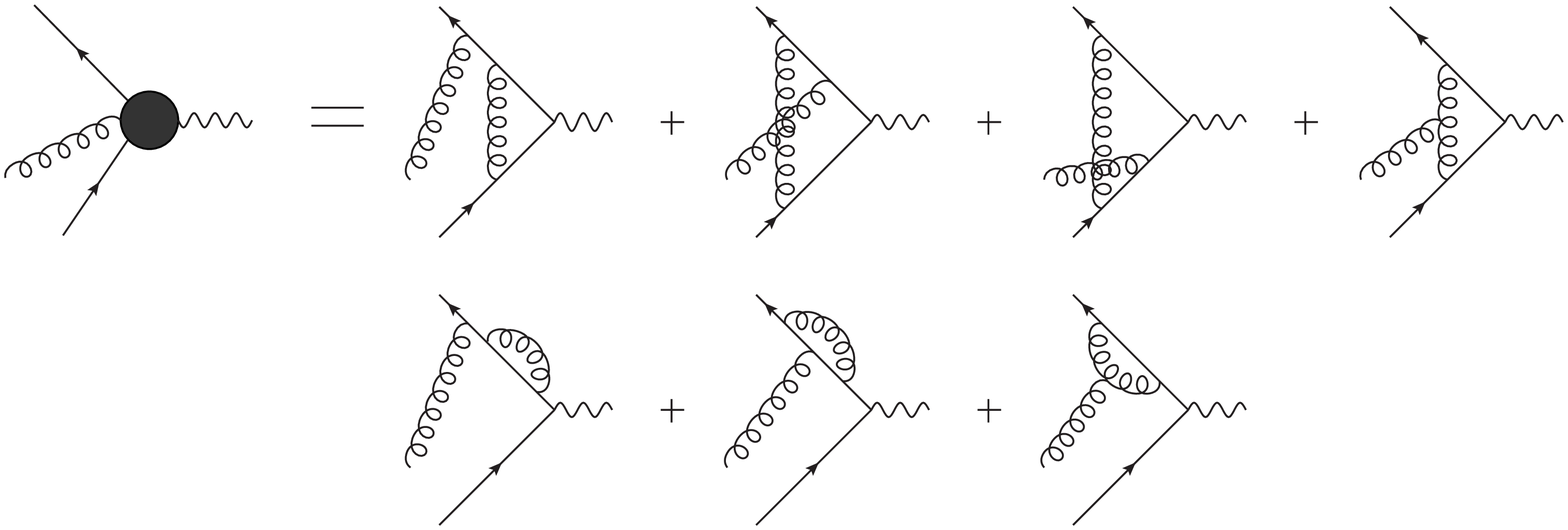, width=5in}
\caption{Virtual corrections to $q$-$\bar q$-$\gamma^*$ vertex with gluon rescattering. }
\label{fig-v-explicit}
\eef

\subsubsection{single-triple scattering interference}

For real corrections to the gluon-quark double scatterings at NLO, besides the central-cut diagrams considered above, there are also asymmetric-cut diagrams, as shown in Fig.~\ref{fig-gq-rc}. They represent the interference between single and triple 
scattering amplitudes, which also have to be included for the complete contribution at the NLO.

In Fig.~\ref{fig-gq-rc}, the short bars (crosses) again stand for the propagators where the soft (hard) poles arise. These poles can be both soft as shown in Fig.~\ref{fig-gq-rc} (left); soft-hard as shown in Fig.~\ref{fig-gq-rc} (middle); or hard-soft as shown in Fig.~\ref{fig-gq-rc} (right), respectively. Of course one also has the interference diagrams where two exchanged gluons on the right side of the cut. We perform the calculations similarly as above, and find several interesting facts. First, for soft-soft double scattering, the asymmetric-cut diagrams only lead to a ``contact'' contribution (thus no nuclear-size enhancement and can be neglected), similar to the case as demonstrated for SIDIS process previously~\cite{Luo:1994np,Kang:2014ela}. On the other hand, the soft-hard or hard-soft asymmetric-cut diagrams have no divergences but only finite NLO corrections. We collect the explicit expressions for these finite NLO corrections in the Appendix: $T_{gq}\otimes H_{gq-R}^{sh}$ and $T_{gq}\otimes H_{gq-R}^{hs}$ in Eqs.~\eqref{eq-gq-R-sh} and \eqref{eq-gq-R-hs} stand for the finite terms of soft-hard and hard-soft rescatterings, respectively, with subscript ``R'' representing the diagrams where the cut is on the right sdie of the two exchanged gluons. Similarly, $T_{gq}\otimes H_{gq-L}^{sh}$ and $T_{gq}\otimes H_{gq-L}^{hs}$ in Eqs.~\eqref{eq-gq-L-sh} and \eqref{eq-gq-L-hs} represent for contributions from those diagrams where the cut is on the left side of the two exchanged gluons. 
\bef
\psfig{file=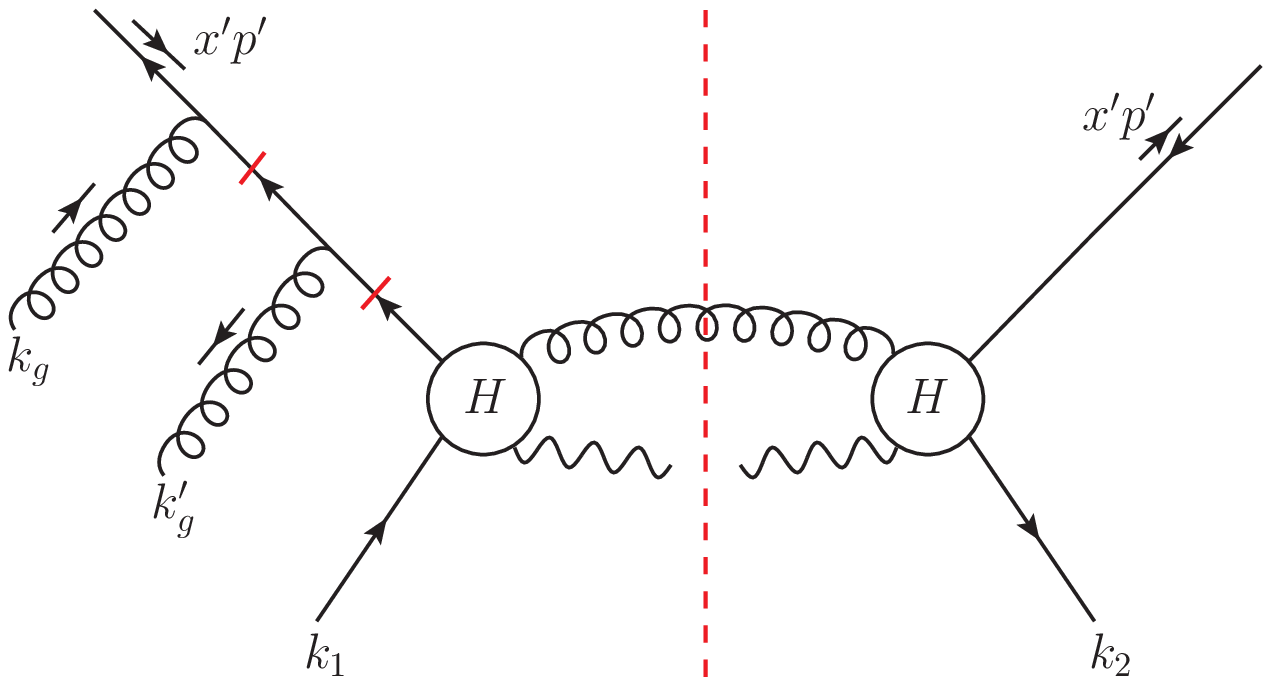, width=2.23in}
\hskip 0.1in 
\psfig{file=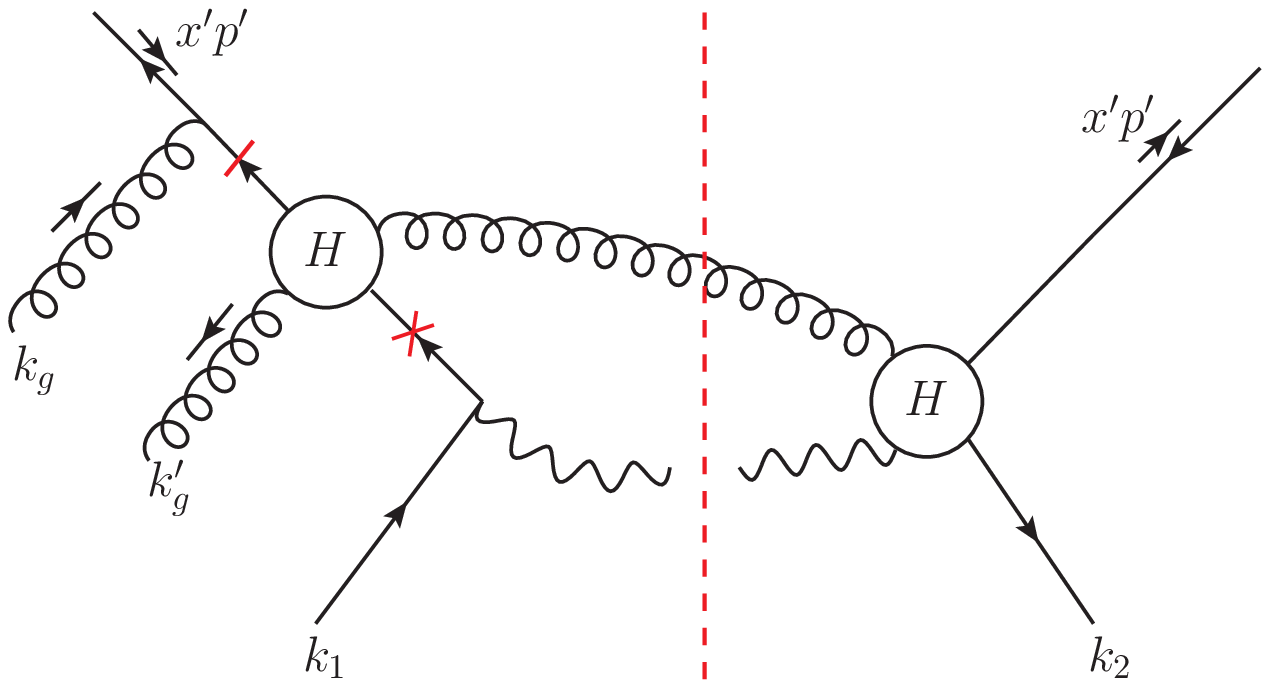, width=2.23in}
\hskip 0.1in 
\psfig{file=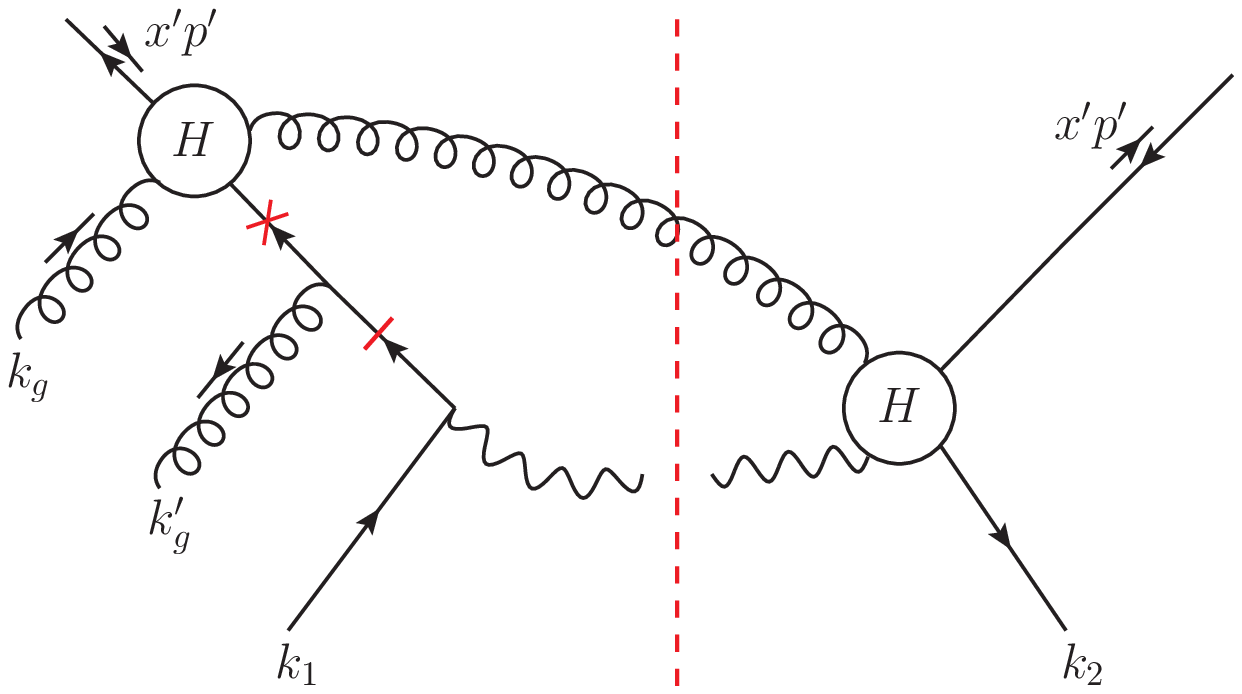, width=2.23in}
\caption{Interference diagrams between triple and single scatterings, for soft-soft (left), soft-hard (middle) and hard-soft (right) rescattering cases, respectively. }
\label{fig-gq-rc}
\eef

\subsubsection{final result from gluon-quark double scattering}

We will now combine all the contributions from the gluon-quark double scattering, which include both real and virtual diagrams. For real contributions, the divergent terms only come from the central-cut diagrams with soft-soft, hard-hard, soft-hard and hard-soft double scatterings, and they are given by
\bea
\frac{d\langle q_T^2\sigma^D\rangle^{\rm (R)}}{dQ^2} \stackrel{\rm div}{=} & \sigma_{\ell} \frac{\alpha_s}{2\pi}\sum_q
\int\frac{dx'}{x'}f_{\bar q/p}(x')
\int_{x_B}^{1}\frac{dx}{x}\int_0^1dv\left(\frac{4\pi\mu^2}{Q^2}\right)^{\epsilon}
\frac{1}{\Gamma(1-\epsilon)}
\Bigg\{\frac{1}{\epsilon^2}\delta(1-z)\left[\delta(1-v)+\delta(v)\right]T_{gq}(x,0,0)C_F
\nnu
&-\frac{1}{\epsilon}\delta(1-v)T_{gq}(x,0,0)C_F\frac{1+z^2}{(1-z)_+}
-\frac{1}{\epsilon}\delta(v)\bigg[T_{gq}(x,0,0)C_F\frac{1+z^2}{(1-z)_+}
+T_{gq}(xz,x(1-z),0)C_A\frac{2}{(1-z)_+}
\nnu
&
-\big(T_{gq}\left(x,0,-x(1-z)\right)+T_{gq}\left(xz,x(1-z),x(1-z)\right)\big)\frac{C_A}{2}\frac{1+z}{(1-z)_+}
+T_{gq}(x,0,0)C_A2\delta(1-z)\bigg]\Bigg\},
\label{eq-real}
\eea
where ``div'' above the equal sign indicates that only the divergent terms are written down on the right hand side of the equation. As one can see, we have both soft and collinear divergences, as $\epsilon \to 0$.  The $1/\epsilon^2$ term is responsible for the soft divergence in the collinear region, while the terms proportional to $1/\epsilon$ are caused by either the soft or collinear divergence, but, not both. On the other hand, the virtual contribution is given by Eq.~\eqref{eq:virtual}. By combining both real and virtual contributions, we find that all double pole terms cancel out as required by the collinear factorization. The single pole term of the virtual contribution, proportional to $-3/\epsilon\, \delta(1-z)$, should also cancel 
the soft (or infrared) divergent part of the $1/\epsilon$ terms of the real contribution in Eq.~(\ref{eq-real}), so that we are left with only the collinear divergent single pole terms. The final result for gluon-quark double scattering has the following expression
\bea
\frac{d\langle q_T^2\sigma^D\rangle_{gq}}{dQ^2} =& \sigma_{\ell} \frac{\alpha_s}{2\pi}\sum_q
\int\frac{dx'}{x'}f_{\bar q/p}(x')
\int_{x_B}^{1}\frac{dx}{x}\int_0^1dv \left(\frac{4\pi\mu^2}{Q^2}\right)^{\epsilon}
\frac{1}{\Gamma(1-\epsilon)}
\bigg[-\frac{1}{\epsilon}\delta(1-v)T_{gq}(x, 0, 0) P_{qq}(z)
\nnu
&
-\frac{1}{\epsilon}\delta(v)T_{gq}\otimes {\mathcal P}_{qg\to qg}
+{\rm finite ~terms}
\bigg],
\label{eq-gq-div}
\eea
where we have suppressed the finite terms for now, the subscript ``gq'' represents the gluon-quark double scattering, and $P_{qq}$ is the normal $q\to q$ splitting function given by
\bea
P_{qq}=C_F\left[\frac{1+z^2}{(1-z)_+}+\frac{3}{2}\delta(1-z)\right].
\eea
On the other hand, the short-hand notation $T_{gq}  \otimes {\mathcal P}_{qg\to qg}$ has the following form
\bea
T_{gq}  \otimes {\mathcal P}_{qg\to qg} \equiv  &
P_{qq}(z) T_{gq}(x, 0, 0) 
+ \frac{C_A}{2} \bigg[ \frac{4}{(1-z)_+} 
T_{qg}(xz, x(1-z), 0) - \frac{1+z}{(1-z)_+}
\big(T_{qg}(x,0,-x(1-z))
\nnu
&
+T_{qg}(xz,x(1-z),x(1-z))
\big)\bigg]
+2C_A\delta(1-z)T_{gq}(x,0,0).
\eea
As discussed in next section, the two collinear divergent $1/\epsilon$ terms, as $\epsilon\to 0$ in Eq.~(\ref{eq-gq-div}), will be absorbed into the normal twist-2 antiquark distribution of the colliding proton and the twist-4 gluon-quark correlation function of the colliding heavy ion, respectively, as required by the QCD collinear factorization at the next leading power accuracy.

\subsection{gluon-gluon double scattering}

In this subsection, we consider gluon-gluon double scattering in DY production at NLO. The relevant central-cut Feynman diagrams are shown in Fig.~\ref{fig-NLO-gg}, with the ``H''-blob given by Fig.~\ref{fig-qg2photonq}. 
\bef
\psfig{file=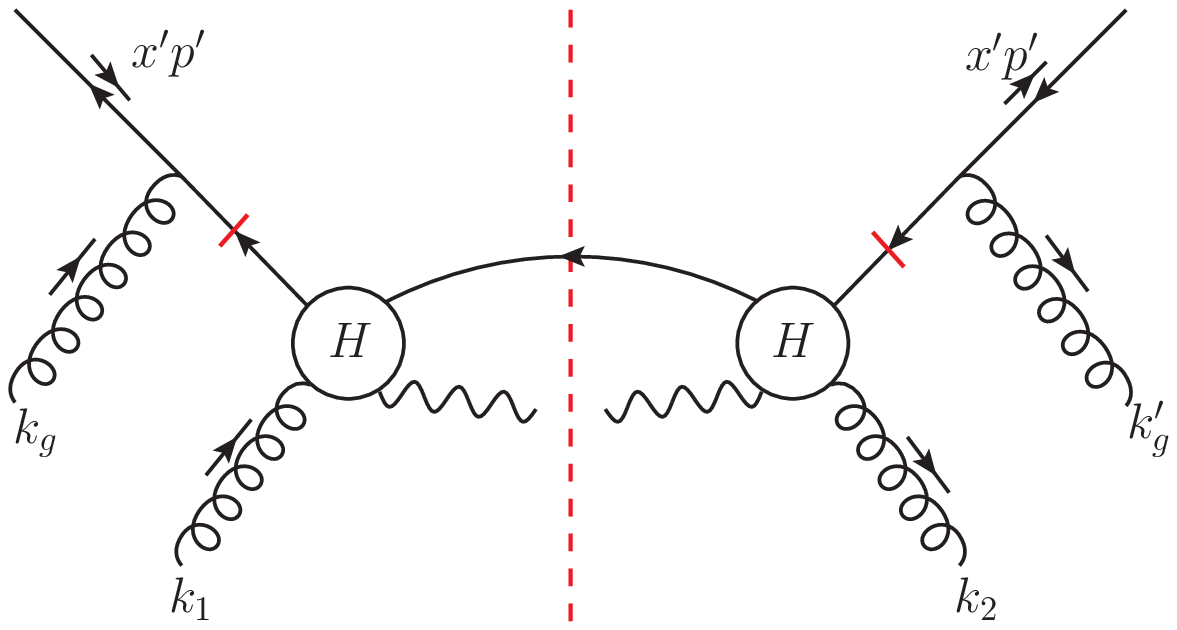, width=2.5in}
\caption{The central-cut diagram for soft-soft gluon-gluon double scattering in DY process. The ``H''-blob represents the hard $2\to 2$ process ($\bar q+g\to \gamma^*+\bar q$) as shown in Fig. \ref{fig-qg2photonq}.}
\label{fig-NLO-gg}
\eef

\bef
\psfig{file=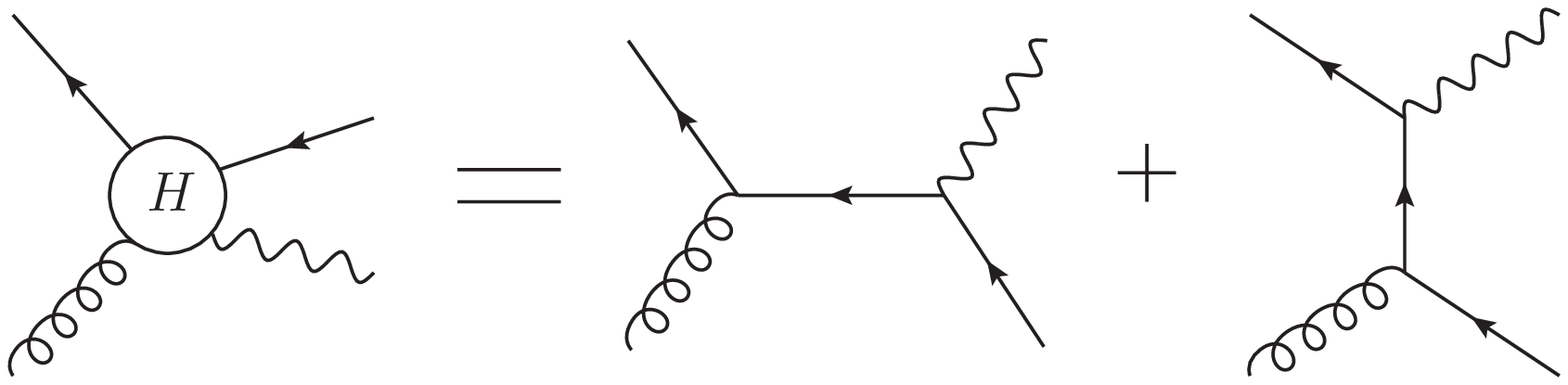, width=3.7in}
\caption{The representation of hard process for $\bar q+g\to \gamma^*+\bar q$.}
\label{fig-qg2photonq}
\eef
Gluon-gluon double scattering receives only soft-soft rescattering contributions, and thus the short bars in Fig.~\ref{fig-NLO-gg} represent the propagators where the soft poles arise. Such a double scattering happens as follows: first $\bar q+g\to \bar q$ through a soft gluon exchange, and then a $2\to 2$ Compton hard interaction $\bar q+g\to \gamma^*+ \bar q$. Being a soft-soft contribution, we follow the same method as in soft-soft gluon-quark double scattering as discussed in the last subsection, and obtain the following result:
\bea
\frac{d\langle q_T^2\sigma^D\rangle_{gg}}{dQ^2} &= \sigma_{\ell} \frac{\alpha_s}{2\pi}
\sum_q \int\frac{dx'}{x'}f_{\bar q/p}(x')
\int_{x_B}^{1}\frac{dx}{x}\int_0^1dv \left(\frac{4\pi\mu^2}{Q^2}\right)^{\epsilon}
\frac{1}{\Gamma(1-\epsilon)}
\left[-\frac{1}{\epsilon}\delta(v)T_{gg}(x,0,0)P_{qg}(z)
+T_{gg}\otimes H_{gg-C}^{ss}
\right],
\label{eq-gg-div}
\eea
where the subscript ``gg'' stands for the gluon-gluon double scattering contribution, with the gluon-gluon correlation function $T_{gg}$ given by the following operator definition:
\bea
T_{gg}(x, 0, 0) =
\frac{1}{xp^+}\int \frac{dy^-}{2\pi} e^{ixp^+y^-}  \int \frac{dy_1^-dy_2^-}{2\pi} 
\theta(-y_2^-)\theta(y^- - y_1^-)
\langle A| F_{\sigma}^+(y_2^-) F_{\alpha}^+(0)F^{+\alpha}(y^-) F^{\sigma +}(y_1^-) |A\rangle.
\label{Tgg}
\eea

The expression for the finite term $T_{gg}\otimes H_{gg-C}^{ss}$ in Eq.~\eqref{eq-gg-div} is collected in Eq.~\eqref{eq-gg-ss} in the Appendix. On the other hand, the divergent piece is proportional to $P_{qg}$, the usual $g\to q$ splitting function,
\bea
P_{qg}(z)=T_R\left[z^2+\left(1-z\right)^2\right],
\eea
with $T_R=1/2$ the color factor. It might be instructive to mention that there is only a collinear divergence for gluon-gluon double scattering at this order. Such a divergence comes from the phase space where the splitted antiquark $\bar q$ of the right diagram in Fig.~\ref{fig-qg2photonq} is radiated collinear to the incoming gluon of the colliding nucleus, as signaled by $\delta(v)$ in Eq.~\eqref{eq-gg-div}. 
This collinear divergence should be absorbed into the gluon-quark correlation functions, $T_{gq}$, due to the one-loop evolution from the gluon-gluon correlation function $T_{gg}$ to the gluon-quark correlation function $T_{gq}$.

In addition to the central-cut diagram as shown in Fig.~\ref{fig-NLO-gg}, one should also take into account the interference between single and triple scattering amplitudes, i.e. asymmetric-cut diagrams where two exchanged gluons are on the same side of the cut. However, similar to the soft-soft gluon-quark double scattering above, all of the asymmetric-cut diagrams lead only to ``contact'' contributions, which can be neglected due to the lack of nuclear enhancement. 

\subsection{Final result and QCD evolution equation for gluon-quark correlation function}

We are now ready to combine the NLO perturbative corrections to DY transverse momentum broadening for both gluon-quark and gluon-gluon double scattering contributions. By adding the expressions in Eqs.~\eqref{eq-gq-div} and \eqref{eq-gg-div}, we obtain the transverse momentum weighted DY cross section from double scattering in p+A collisions as follows:
\bea
\frac{d\langle q_T^2\sigma^D\rangle}{dQ^2} =& \sigma_{\ell} \frac{\alpha_s}{2\pi}\sum_q
\int\frac{dx'}{x'}f_{\bar q/p}(x')
\int_{x_B}^{1}\frac{dx}{x}\int_0^1dv
\Bigg\{\left(-\frac{1}{\hat \epsilon}+\ln\frac{Q^2}{\mu^2}\right)
\bigg[\delta(1-v)T_{gq}(x, 0, 0)P_{qq}(z)
\nnu
&
+\delta(v)\Big(T_{gq}\otimes {\mathcal P}_{qg\to qg}+T_{gg}(x,0,0)P_{qg}(z)\Big)\bigg]
+{\rm finite ~terms}
\Bigg\},
\label{eq-div}
\eea
where $1/\hat \epsilon = 1/\epsilon - \gamma_E + \ln 4\pi$. The finite terms are collected in the Appendix, and will be specified below. We will first discuss the divergent pieces.

There are two types of collinear divergences in Eq.~\eqref{eq-div}, which are associated with $\delta(1-v)$ and $\delta(v)$, respectively. As we have discussed already in Sec.~\ref{ss-gq}, $v=1$ corresponds to the situation where the radiated gluon is collinear to the beam proton, and thus such a collinear divergence should be absorbed into the QCD evolved parton distribution functions of the projectile proton. 
At the order of $\alpha_s$ in the $\overline{\rm MS}$-scheme, we have
\bea
f_{\bar q/p}(x, \mu_f^2) =  f_{\bar q/p}(x) - \frac{\alpha_s}{2\pi} \left(\frac{1}{\hat\epsilon}+\ln\frac{\mu^2}{\mu_f^2}\right)\int \frac{dx'}{x'}  f_{\bar q/p}(x') P_{qq}(z),
\eea
where $z=x/x'$, $\mu$ and $\mu_f$ are the renormalization and factorization scale, respectively, 
and $f_{\bar q/p}(x)$ denote the bare LO antiquark distribution in the proton. 
On the other hand, $v\to 0$ (or $\theta\to \pi$) represents the situation where the radiated gluon 
(or antiquark in the case of gluon-gluon double scattering) 
is collinear to the colliding nucleus, and the associated collinear divergence should thus be absorbed into the 
QCD evolved twist-4 gluon-quark correlation function as follows: 
\bea
T_{gq}(x_B, 0, 0, \mu_f^2)  =  T_{gq}(x_B,0,0) -\frac{\alpha_s}{2\pi}\left(\frac{1}{\hat\epsilon}+\ln\frac{\mu^2}{\mu_f^2}\right)\int_{x_B}^1\frac{dx}{x}
\Big[{\mathcal P}_{qg\to qg} \otimes T_{gq}+P_{qg}(z)T_{gg}(x,0,0)\Big],
\label{eq-redefine}
\eea 
where $T_{gq}(x_B,0,0)$ is the bare LO gluon-quark correlation function. From Eq.~\eqref{eq-redefine}, one could derive the following QCD evolution equation for the gluon-quark correlation function $T_{gq}$:
\bea
\frac{\partial}{\partial \ln \mu_f^2} T_{gq}(x_B,0,0,\mu_f^2)  =  \frac{\alpha_s}{2\pi} 
\int_{x_B}^1 \frac{dx}{x} \Big[{\mathcal P}_{qg\to qg} \otimes T_{qg} 
+ P_{qg}(z) T_{gg}(x, 0, 0, \mu_f^2)\Big],
\label{eq-evolution}
\eea
which could also get contributions from other twist-4 parton-parton correlation functions, such as quark-quark correlation functions, $T_{qq}$, or quark-antiquark correlation functions, etc.  We will come back to this in next section.

Before concluding this section, it is important to point out that the QCD evolved gluon-quark correlation function at NLO as given in Eq.~\eqref{eq-redefine} for DY process (representing initial-state double scattering) has the same form as the corresponding quark-gluon correlation function at NLO for SIDIS process (representing final-state double scattering) as given in our previous paper~\cite{Kang:2013raa,Kang:2014ela}~\footnote{In the operator definitions, the only difference lies in the different $\theta$-functions, with $\theta(-y_2^-)\theta(y^- - y_1^-)$ for DY while $\theta(y_2^-)\theta(y_1^- - y^-)$ for SIDIS, which are nothing but simply representing the order of double scatterings.}. The QCD evolution equation for the gluon-quark correlation function derived in DY process is also the same as that for the quark-gluon correlation function derived in SIDIS process. This confirms the collinear factorization for the transverse momentum broadening at the NLO, and demonstrates the universality of the associated gluon-quark correlation function. This in turn indicates that 
the properties of nuclear matter (contained in this twist-4 correlation function) as probed by an energetic parton are independent of the hard process that creates the energetic parton. Therefore, one can ``measure'' this correlation funciton in one process (e.g. SIDIS) and then apply it to some other processes (e.g. DY), which is consistent with the QCD factorization at twist-4 and can be regarded as the prediction power of perturbative QCD for parton multiple scattering in nuclear medium.

After absorbing all the collinear divergences into the QCD evolved nonperturbative distribution or correlation functions, we obtain the transverse momentum weighted differential cross section for DY process in p+A collisions at twist-4 up to NLO accuracy:
\bea
\frac{d\langle q_T^2\sigma^D\rangle}{dQ^2} =& 
\sigma_{\ell} \sum_q
\int\frac{dx'}{x'}f_{\bar q/p}(x',\mu_f^2)
\Bigg\{
\int_{x_B}^{1}\frac{dx}{x}T_{gq}(x,0,0,\mu_f^2)\delta(1-z)\nnu
&+\frac{\alpha_s}{2\pi} 
\int_{x_B}^{1}\frac{dx}{x}
\bigg[\ln\left(\frac{Q^2}{\mu_f^2}\right)
P_{qq}(z) T_{gq}(x,0,0,\mu^2)
\nnu
&
+ \ln\left(\frac{Q^2}{\mu_f^2}\right) 
\Big(T_{gq}\otimes {\mathcal P}_{qg\to qg}+P_{qg}(z)T_{gg}(x,0,0,\mu_f^2)\Big)
+ H^{\rm NLO} \bigg]
\Bigg\},
\label{eq-NLO}
\eea
where $x_B=Q^2/2p\cdot q$ is the Bjorken variable for DY process, $z=x_B/x$, and $H^{\rm NLO}$ represents finite terms and has the following expression
\bea
H^{\rm NLO} =& T_{gq}\otimes H^{ss}_{gq-C}+T_{gq}\otimes H^{hh}_{gq-C}+T_{gq}\otimes H^{sh}_{gq-C}+T_{gq}\otimes H^{hs}_{gq-C}
\nnu
& -T_{gq}\otimes H^{sh}_{gq-R}-T_{gq}\otimes H^{hs}_{gq-R}-T_{gq}\otimes H^{sh}_{gq-L}-T_{gq}\otimes H^{hs}_{gq-L}
\nnu
& +T_{gq}\otimes H^{V}_{gg-C}+T_{gg}\otimes H^{ss}_{gg-C}.
\label{eq:HNLO}
\eea
Here the first line in Eq.~\eqref{eq:HNLO} stands for the contributions from the central-cut real diagrams, and the second line are the contributions from asymmetric-cut diagrams, i.e. the interference between single and triple scattering amplitudes. The first term in the last line is the virtual correction, and the second term in the last line is the contribution from the gluon-gluon double scattering. Except the virtual correction $T_{gq}\otimes H^{V}_{gg-C}$ that is given by Eq.~\eqref{eq:virtual-finite}, the explicit expressions for all the other terms in Eq.~\eqref{eq:HNLO} are collected in the Appendix. Eq.~\eqref{eq-NLO} is the main result of our paper. By dividing this $q_T^2$-weighted DY cross section from double scattering in Eq.~\eqref{eq-NLO} by the NLO inclusive DY cross section, as in the definition in Eq.~\eqref{eq-weight}, one can then compute the transverse momentum broadening $\Delta\langle q_T^2\rangle$ for DY process in p+A collisions at the NLO accuracy. 

\section{summary}
\label{sec-sum}
In this paper, we calculate the NLO contributions to the nuclear transverse momentum broadening for Drell-Yan dilepton production in p+A collisions. Specifically, we evaluated the transverse-momentum-weighted differential cross section at twist-4 at NLO, by taking into account all the contributions from both gluon-quark and gluon-gluon double scatterings, as well as the interference between single and triple scattering amplitudes.  Through explicit calculations, we demonstrate that the soft divergences cancel out between real and virtual corrections. The remaining collinear divergences can be absorbed into 
either the QCD evolved parton distribution function of the beam proton, or the QCD evolved twist-4 nuclear gluon-quark correlation function of the colliding nucleus, which leads to a DGLAP-type evolution equation for the twist-4 gluon-quark correlation function, same as that derived in the NLO calculation in SIDIS $e$+A collisions. This verifies the universality of the 
twist-4 gluon-quark correlation function, and in turn indicates that the properties of nuclear matter as probed by a hard probe (scattering with a large momentum transfer) are independent of the the details of the hard process. Such a property also demonstrates the prediction power of perturbative QCD approach for parton multiple scattering in nuclear medium. After the subtraction of all the collinear divergences, the transverse-momentum-weighted DY cross section can be factorized into the convolution of a normal twist-2 parton distribution functions of the colliding proton, convoluted with a gluon-quark (or gluon-gluon) correlation function of the colliding nucleus and the perturbatively calculated hard coefficient functions accurate up to one loop order. We expect this improved NLO result for DY transverse momentum broadening in p+A collisions can be applied phenomenologically to compare with data from ongoing experiments at FNAL, RHIC and LHC, to test QCD treatment of partonic multiple scatterings beyond the existing LO accuracy.  

For the complete NLO contribution to the DY transverse momentum broadening from partonic double scatterings, we also need to calculate the contributions from (anti)quark-(anti)quark double scattering subprocesses, which will also have contributions from the soft-soft, hard-hard, soft-hard, and hard-soft double scattering channels.  Unlike the gluon-quark and gluon-gluon double scattering subprocesses, the collinear expansion for (anti)quark-(anti)quark double scattering is more straightforward, because there is no need for the complication to convert the gluon field operators into field strength operators in order to define the good twist-4 gluon-quark and gluon-gluon correlation functions.  We will present our calculations of full contributions to the DY transverse momentum broadening from (anti)quark-(anti)quark double scattering subprocesses in a future publication.

\section*{Acknowledgments}
This work is supported by the U.S. Department of Energy under Contract Nos.~DE-AC52-06NA25396 (Z.K. and H.X.), DE-AC02-98CH10886 (J.Q.) and DE-AC02-05CH11231 (X.W.), and in part by the National Science Foundation of China under Grants Nos.~11221504 and 10825523, China Ministry of Science and Technology under Grant No.~2014DFG02050, and the Major State Basic Research Development Program in China (No.~2014CB845404).

\appendix*
\section{complete list of finite terms}
In this appendix, we list the finite terms for the transverse momentum weighted cross section in DY process from gluon-quark and gluon-gluon double scatterings. For central-cut diagrams in gluon-quark double scattering, the finite pieces for soft-soft, hard-hard, soft-hard and hard-hard double scatterings are denoted as $T_{gq}\otimes H_{gq-C}^{ss}$, $T_{gq}\otimes H_{gq-C}^{hh}$, $T_{gq}\otimes H_{gq-C}^{sh}$ and $T_{gq}\otimes H_{gq-C}^{hs}$, respectively. For asymmetric-cut diagrams, we use $T_{gq}\otimes H_{gq-R}^{sh}$ and $T_{gq}\otimes H_{gq-R}^{hs}$ to represent the finite term for soft-hard and hard-soft rescatterings, respectively.  Similarly, $T_{gq}\otimes H_{gq-L}^{sh}$ and $T_{gq}\otimes H_{gq-L}^{hs}$ stands for the finite term for soft-hard and hard-soft rescatterings, respectively. Here the subscript ``C'' represents the central-cut diagrams, ``R'' (``L'') stands for the diagrams where the cut is on the right (left) side of the two exchanged gluons. Finally, the finite term for gluon-gluon double scattering is denoted by $T_{gg}\otimes H_{gg}^{ss}$, which has only soft-soft double scattering contributions. These finite terms are given by the following expressions:
\bea
T_{gq}\otimes H_{gq-C}^{ss}=&x^2\frac{d^2T_{gq}(x,0,0)}{dx^2}C_F\bigg[
\frac{1}{3}(5z^3-3z^2+3z-5)+\ln\frac{(1-z)^2}{z}(1-z)(1+z^2)\bigg]
+x\frac{dT_{gq}(x,0,0)}{dx}C_F
\nnu
&\times
\bigg[
\frac{1}{3}(-20z^3+18z^2+6z+8)-\ln\frac{(1-z)^2}{z}(-4z^3+5z^2+1)
\bigg]
+T_{gq}(x,0,0)C_F\bigg\{
4\delta(1-z)
\nnu
&
-\frac{z^4+4z^3-6z^2+8z+1}{(1-z)_+}
-\frac{1}{3}(1-z)(32z^2-7z+5)
+\bigg[
2\bigg(\frac{\ln(1-z)}{1-z}\bigg)_+-\frac{\ln z}{(1-z)_+}
\bigg]
\nnu
&\times
(7z^4-14z^3+10z^2+1)
\bigg\},
\label{eq-gq-ss}
\\
T_{gq}\otimes H_{gq-C}^{hh}=&\bigg[
2\bigg(\frac{\ln(1-z)}{1-z}\bigg)_+-\frac{\ln z}{(1-z)_+}
\bigg]\bigg\{2C_AT_{gq}(xz,x(1-z),0)+(1+z^2)\big[C_Az+C_F(1-z)^2\big]T_{gq}(x,0,0)\bigg\}\nnu
&
+(1-z)\big[C_Az+C_F(1-z)^2\big]T_{gq}(x,0,0)
+\int_0^1 dv \,T_{gq}(x_d,x-x_d,0)
\nnu
&\times
\frac{\big[(1-v+vz)^2+1\big]\big[C_A(vz-v+1)+C_Fv^2(1-z)^2\big]}{(1-z)_+v_+(1-v)_+},
\label{eq-gq-hh}
\\
T_{gq}\otimes H_{gq-C}^{sh}=&\int_0^1dv\frac{\big[(1-v+vz)^3+z\big]\big[C_A+2C_Fv(z-1)\big]}
{2v_+(1-v+vz)}
x\frac{d}{dx}T_{gq}(x,0,x_d-x)\nnu
&
+\frac{1}{2}C_A(1+z)\bigg[\ln\frac{(1-z)^2}{z}-1\bigg]
x\frac{d}{dx}T_{gq}\big[x,0,-x(1-z)\big]
-\delta(1-z)2C_AT_{gq}(x,0,0)
\nnu
&-
\int_0^1dv\frac{\big[(1-v+vz)^3+z\big]\big[C_A+2C_Fv(z-1)\big]}
{2(1-v+vz)^2}(1-z)
x\frac{d}{dx_d}T_{gq}(x,0,x_d-x)\nnu
&
-\frac{1}{2(1-z)_+}
\bigg\{C_A(z^2-3z-2)T_{gq}\big[x,0,-x(1-z)\big]
+\big[C_A+2C_F(z-1)\big](z^3-4z^2+z-2)\nnu
&\times
T_{gq}(x,0,0)\bigg\}
+\bigg[
\bigg(\frac{\ln(1-z)}{1-z}\bigg)_+-\frac{\ln z}{2(1-z)_+}
\bigg]\bigg\{C_A(z^2-2z+1)T_{gq}\big[x,0,-x(1-z)\big]-z(1+z^2)
\nnu
&\times
\big[C_A-2C_F(1-z)\big]T_{gq}(x,0,0)\bigg\}
-\int_0^1dv\frac{C_A+2C_Fv(z-1)}
{2(1-z)_+v_+(1-v)_+(1-v+vz)}
\big[(1-v+vz)^4\nnu
&
+(2v-1)(z-1)z+z\big] T_{gq}(x,0,x_d-x),
\label{eq-gq-sh}
\\
T_{gq}\otimes H_{gq-C}^{hs}=&\int_0^1dv\frac{\big[(1-v+vz)^3+z\big]\big[C_A+2C_Fv(z-1)\big]}
{2v_+(1-v+vz)}
x\frac{d}{dx}T_{gq}(x_d,x-x_d,x-x_d)
\nnu
&+\frac{1}{2}C_A(1+z)\bigg[\ln\frac{(1-z)^2}{z}-1\bigg]
x\frac{d}{dx}T_{gq}\big[xz,x(1-z),x(1-z)\big]
-\delta(1-z)2C_AT_{gq}(x,0,0)
\nnu
&
-\int_0^1dv\frac{\big[(1-v+vz)^3+z\big]\big[C_A+2C_Fv(z-1)\big]}
{2(1-v+vz)^2}(1-z)
x\frac{d}{dx_d}T_{gq}(x_d,x-x_d,x-x_d)
\nnu
&
-\frac{1}{2(1-z)_+}
\bigg\{C_A(z^2-3z-2)T_{gq}\big[xz,x(1-z),x(1-z)\big]
+\big[C_A+2C_F(z-1)\big](z^3-4z^2+z-2)
\nnu
&\times T_{gq}(x,0,0)\bigg\}
+\bigg[
\bigg(\frac{\ln(1-z)}{1-z}\bigg)_+-\frac{\ln z}{2(1-z)_+}
\bigg]\bigg\{C_A(z^2-2z+1)T_{gq}\big[xz,x(1-z),x(1-z)\big]
\nnu
&-z(1+z^2)
\big[C_A+2C_F(z-1)\big]T_{gq}(x,0,0)\bigg\}
-\int_0^1dv\frac{\big[C_A+2C_Fv(z-1)\big]}
{2(1-z)_+v_+(1-v)_+(1-v+vz)}
\nnu
&\times \big[(1-v+vz)^4+(2v-1)(z-1)z+z\big]
T_{gq}(x_d,x-x_d,x-x_d),
\label{eq-gq-hs}
\\
T_{gq}\otimes H_{gq-R}^{sh}=&\int_0^1 dv \frac{\big[(1-v+vz)^3+z\big]\big[C_A+2C_Fv(z-1)\big]}{2(1-v+vz)^2}
(1-z)x\frac{d}{dx_d}T_{gq}^R(x_d,0,x-x_d)
\nnu
& -(1+z^2)\big[C_A+2C_F (z-1)\big]T_{gq}^R(x,0,0),
\label{eq-gq-R-sh}
\\
T_{gq}\otimes H_{gq-R}^{hs}=&-\int_0^1 dv \frac{\big[(1-v+vz)^3+z\big]\big[C_A(1-v+vz)-2C_Fv(z-1)\big]}{2(1-v+vz)^2}
(1-z)
\bigg(x\frac{d}{dx_d}T_{gq}^R(x_d,x-x_d,x-x_d)
\nnu
&+\frac{d}{dx_2}T_{gq}^R(x_d,x_2,x-x_d)|_{x_2\to x-x_d}\bigg),
\label{eq-gq-R-hs}
\\
T_{gq}\otimes H_{gq-L}^{sh}=&\int_0^1 dv \frac{\big[(1-v+vz)^3+z\big]\big[C_A+2C_Fv(z-1)\big]}{2(1-v+vz)^2}
(1-z)x\frac{d}{dx_d}T_{gq}^R(x,x_d-x,x_d-x)
\nnu
&-(1+z^2) \big[C_A+2C_F(z-1)\big]T_{gq}^R(x,0,0),
\label{eq-gq-L-sh}
\\
T_{gq}\otimes H_{gq-L}^{hs}=&-\int_0^1 dv \frac{\big[(1-v+vz)^3+z\big]\big[C_A(1-v+vz)-2C_Fv(z-1)\big]}{2(1-v+vz)^2}
(1-z)
\bigg(x\frac{d}{dx_d}T_{gq}^R(x,0,x_d-x)
\nnu
&+\frac{d}{dx_2}T_{gq}^R(x,x_2,x_d-x)|_{x_2\to 0}\bigg),
\label{eq-gq-L-hs}
\\
T_{gg}\otimes H_{gg-C}^{ss}=&x^2\frac{d^2}{dx^2}T_{gg}(x,0,0)T_R
\bigg[\frac{1}{12}(67z^3-133z^2+83z-17)(1-z)+\ln\frac{(1-z)^2}{z}(1-z)^2(2z^2-2z+1)\bigg]\nnu
&-x\frac{d}{dx}T_{gg}(x,0,0)T_R(1-z)\bigg[\frac{1}{12}(381z^3-605z^2+253z-29)+\ln\frac{(1-z)^2}{z}(1-2z)(6z^2-6z+1)\bigg]
\nnu
&+T_{gg}(x,0,0)T_R(1-z)\bigg[\frac{1}{12}(737z^3-1029z^2+345z-29)+\ln\frac{(1-z)^2}{z}(-24z^3+30z^2-10z+1)\bigg],
\label{eq-gg-ss}
\eea
where the matrix element $T_{gq}^{R}$ and $T_{gq}^{L}$ are given by
\bea
T_{gq}^R(x_1, x_2, x_3)
=&\int \frac{dy^-}{2\pi} e^{ix_1p^+y^-}  \int \frac{dy_1^-dy_2^-}{4\pi} e^{ix_2p^+(y_1^- - y_2^-)}
 e^{ix_3p^+y_2^-}  \theta(y_2^--y_1^-)\theta(y^- - y_2^-)
\nnu
&
\times  \langle A| F_{\sigma}^+(y_2^-) {\bar\psi}_q(0) \gamma^+ \psi_q(y^-) F^{\sigma +}(y_1^-) |A\rangle,
\\
T_{gq}^L(x_1, x_2, x_3)
=&\int \frac{dy^-}{2\pi} e^{ix_1p^+y^-}  \int \frac{dy_1^-dy_2^-}{4\pi} e^{ix_2p^+(y_1^- - y_2^-)}
 e^{ix_3p^+y_2^-} \theta(y_1^--y_2^-)\theta(- y_1^-) 
\nnu
&
\times \langle A| F_{\sigma}^+(y_2^-) {\bar\psi}_q(0) \gamma^+ \psi_q(y^-) F^{\sigma +}(y_1^-) |A\rangle.
\label{Tgq-LR}
\eea

\bibliographystyle{h-physrev5}   
\bibliography{biblio}

\end{document}